\documentclass{article}

\usepackage{arxiv}

\usepackage[utf8]{inputenc} % allow utf-8 input
\usepackage[T1]{fontenc}    % use 8-bit T1 fonts
\usepackage[hidelinks]{hyperref}      % hyperlinks
\usepackage{url}            % simple URL typesetting
\usepackage{booktabs}       % professional-quality tables
\usepackage{amsfonts}       % blackboard math symbols
\usepackage{nicefrac}       % compact symbols for 1/2, etc.
\usepackage{microtype}      % microtypography
\usepackage{lipsum}		% Can be removed after putting your text content
\usepackage{graphicx}
\usepackage{natbib}
\usepackage{doi}
\usepackage{rotating}
\usepackage{subfig}
\usepackage{makecell}
\usepackage{xcolor}
\usepackage[normalem]{ulem}
\usepackage{soul}

\newcommand{\pkg}[1]{{\normalfont\fontseries{b}\selectfont #1}}
\let\proglang=\textsf
\let\code=\texttt

\title{Who are the gatekeepers of economics? Geographic diversity, gender composition, and interlocking editorship of journal boards}

\date{} 					% Or removing it

\author{\href{https://orcid.org/0000-0003-0293-482X}{\includegraphics[scale=0.06]{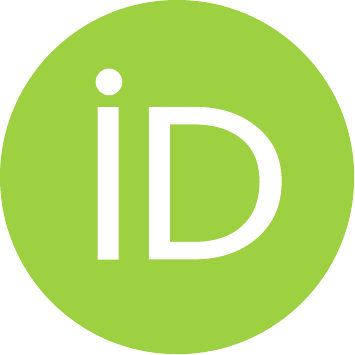}\hspace{1mm}Alberto Baccini}%\thanks{Use footnote for providing further
  \\
	Dipartimento di Economia Politica e Statistica\\
	Università degli Studi di Siena\\
	Siena, Italy \\
	\texttt{alberto.baccini@unisi.it} \\
	\And
	\href{https://orcid.org/0000-0001-7715-8250}{\includegraphics[scale=0.06]{orcid.pdf}\hspace{1mm}Cristina Re} \\
	Dipartimento di Economia Politica e Statistica\\
	Università degli Studi di Siena\\
	Siena, Italy \\
	\texttt{cristina.re@uniupo.it} \\
}

\renewcommand{\headeright}{}
\renewcommand{\undertitle}{}
\renewcommand{\shorttitle}{Who are the gatekeepers of economics?}

\hypersetup{
pdftitle={Who are the gatekeepers of economics?},
pdfauthor={Alberto Baccini, Cristina Re},
}

\begin{document}
\maketitle

\begin{abstract}
This study investigates the role of editorial board members as gatekeepers in science, creating and utilizing a database of 1,516 active economics journals in 2019, which includes more than 44,000 scholars from over 6,000 institutions and 142 countries. The composition of these editorial boards is explored in terms of geographic affiliation, institutional affiliation, and gender. Results highlight that the academic publishing environment is primarily governed by men affiliated with elite universities in the United States. The study further explores social similarities among journals using a network analysis perspective based on interlocking editorship. Comparison of networks generated by all scholars, editorial leaders, and non-editorial leaders reveals significant structural similarities and associations among clusters of journals. These results indicate that links between pairs of journals tend to be redundant, and this can be interpreted in terms of social and intellectual homophily within each board, and between boards of journals belonging to the same cluster. Finally, the analysis of the most central journals and scholars in the networks suggests that journals probably adopt ‘strategic decisions’ in the selection of the editorial board members. The documented high concentration of editorial power poses a serious risk to innovative research in economics.

\end{abstract}

% keywords can be removed

\vfill

\textbf{Acknowledgments:} \textit{This work has been developed as part of a PRIN project 2017MPXW98 funded by the Italian Ministry of University. The project involve the building of a database of the editorial boards of economics journals for the period 1946-2019. The data here presented are only a part of the database and will be released with all others at the end of the project. Thanks to Alberto Montesi for his fundamental bibliographic assistance in finding information about the boards. Eugenio Petrovich contributed to the design of the database and to the development of routines for disambiguation and standardization. Livia Tarini, Martina Cioni and Daria Pignalosa contributed to the building of the database. Thanks to Lucio Barabesi, Oddný Helgadóttir and Jakob Kapeller for their comments and suggestions.}

\section{Introduction}
Quantitative approaches have recently gained increasing attention from economists, as they allow us to uncover aspects of the recent history of economic thought and the professional role of economists that may remain hidden to traditional qualitative methods \citep{duarte2016, marcuzzo2016}. In this paper, we use quantitative tools to investigate the characteristics of gatekeepers in the field of economics. Specifically, we focus on members of editorial boards and editorial leaders of economics journals, as these scholars play a pivotal role in shaping both the trajectory of economic sciences and the careers of economists. Editorial board members are gatekeepers of science \citep{degrazia1963, crane1967}: through their selection of manuscripts to be published in journals, they can influence the direction of research within a discipline by deciding which studies to support and which to reject. They also wield considerable influence over the careers of scholars who seek to publish their work. Given the crucial role that editors play, numerous studies have examined the composition of editorial boards, its correlation with publication outcomes, and its evolution over time.

This paper aims to enhance our understanding of the composition of editorial boards in the field of economics, shedding light on the characteristics of economics gatekeepers. Until now, studies of economics editorial boards have typically employed limited datasets or specific approaches. For instance, \citet{hodgson1999} examined the institutional backgrounds of editors and authors for the top 30 economics journals in 1995, revealing that 70.8\% of journal editors were affiliated with institutions in the United States, with twelve universities accounting for more than 38.9\% of all editors. Their main concern with such a high concentration of institutional power is the threat to ``the potential for innovation and change'' (p.166). A similar concentration of editors affiliated with prestigious institutions was found by \citet{gibbons1991} in a study of the 25 top economics journals from 1970 to 1979.They also
discovered that, among the 575 editors, Harvard had the most members
(36, which corresponds to 9.1 percent of all members), Stanford was second (29 members, 7.3 percent), followed by MIT (25), Chicago (24), and Pennsylvania (22). \citet{wu2020} also noted that academic journals in economics remain heavily dominated by US institutions, with 48.55\% of editors coming from the US, using a sample from 2019 that included 6,916 editors affiliated with 246 economics journals. \citet{addis2003} focused on gender distribution, analyzing the presence of male and female economists on the editorial boards of thirty-six Italian economics journals published from 1970 to 1996. Their findings indicated that women were underrepresented and predominantly occupied lower-ranking positions. \citet{baccini2010} were the first to propose and analyze the interlocking editorship network generated by individuals serving on the editorial boards of multiple economics journals, revealing a cohesive network of editors (90\% of the journals are directly or indirectly connected) containing different components. 
Lastly, \citet{ductor2023} conducted a study involving 106 economics journals spanning the period from 1990 to 2011, employing also IE analysis. Their research underscored the presence of a discipline characterized by a significant concentration of both institutional and individual power, particularly within the more prestigious journals. Furthermore, they emphasized a strong negative correlation between the duration of editorial tenure and the impact of a journal.

This study investigates the country, institutional, and gender distribution of editorial boards in economics journals, as well as the characteristics of the interlocking editorship networks they create, on an unprecedented large scale. Our analysis is based on a comprehensive database that includes all 1,516 journals listed in the \textit{EconLit} database with an active editorial board in 2019. For each journal, we manually compiled a database containing the names of board members and their affiliations, resulting in a dataset with over 44,000 members representing more than 6,000 institutions and 142 countries. This dataset offers an unprecedented opportunity to investigate the phenomenon of gatekeeping in contemporary economics on a large scale.

The article is organized as follows: Section 2 presents a literature review of studies on editorial boards in various fields. In Section 3, the dataset and research questions are described. Sections 4, 5, and 6 report on the geographic distribution, institutional distribution, and gender composition of editorial board members, respectively. These sections outline the differences between all editorial roles and editorial leaders, considering all journals and each journal separately.
%with a specific focus on the top five journals. 
In Section 7, the analysis of the interlocking editorship is presented. Section 8 examines the most central journals and editorial leaders within the interlocking editorship network, separately analyzing the network formed by female scholars. The analysis concludes with some policy recommendations for implementing practices aimed at diversifying the members of editorial boards.

\section{Literature review}
\label{sec:review}

Since the inception of gatekeeping analysis in the sociology of science, significant attention has been directed towards the role of journal editors, who are regarded as the primary gatekeepers of scientific knowledge \citep{degrazia1963, crane1967}. This emphasis on editors likely stems from their pivotal role in shaping the trajectory of scientific knowledge by selecting works deemed worthy of publication. Their activities also indirectly impact the careers of scholars, particularly in the last 20-30 years, as academic success increasingly relies on quantitative bibliometric indicators.
According to \citet{merton1942}, the fundamental role of editors should align with the normative ideal of `universalism,' wherein scientific contributions are evaluated solely based on their intellectual merit. Nevertheless, concerns have arisen regarding the extent to which editors actually promote the best scientific output. These concerns are rooted in worries that social biases, linked to scholars' demographic or institutional characteristics, may also come into play. 
\citet{crane1967} provided empirical evidence that authors' academic affiliations, doctoral origins, and professional age tended to be similar to the distribution of those characteristics among journal editors, and these factors significantly influence editorial decisions in the selection of journal articles. Other studies proved that a narrow composition of the editorial board, in terms of similar education, research background, and academic experience, can restrict the themes and methodologies that are published in a journal (for a comprehensive review, see \citet{mazov2016}). 

For these reasons, numerous studies have centered their focus on the composition of editorial boards, examining its correlation with publication outcomes and its evolution over time. Additionally, investigations into the composition of editorial boards have been employed to evaluate journal internationalization and gender balance. They have also served as indicators of research influence across geographic regions, institutions, gender, and groups of scholars.

In particular, \citet{zsindely1982}, in their examination of the \emph{geographic distribution} of editorial boards across 252 scientific journals, identified a significant correlation between the number of editorial board members from a particular country and the quantity of journals and authors associated with that country. Notably, Israel, Western Europe, the United States, and Canada exhibited an overrepresentation on editorial boards in comparison to their share of academic publications and scholarly journals. Conversely, Japan, India, and the Soviet Union were found to be underrepresented. Larger-scale studies have arrived at similar findings, highlighting that manuscripts submitted by authors from countries outside those of the editorial board members are more likely to face rejection. Additionally, in the case of most international journals, the majority of editorial board members are U.S. citizens (see \citet{mazov2016} for a review). \citet{braun2005a, braun2005b} interpreted this phenomenon as an indication that the United States had held a dominant scientific position since 1982 and that this dominance had not waned up to that point, despite other countries increasing their numbers of published papers and citations. \citet{leydesdorff2009} demonstrated that China had recently become the second-largest nation in terms of both publications and citations, yet this diversification of the research landscape had not yet been reflected in the composition of editorial boards. According to \citet[p.1548]{braun2005a} ``journal papers and citations are just a corollary'' and ``the control and screening activity of journal editorial boards {[}..{]} is of paramount importance''. They believed that the predominance of U.S. scientists as editorial board members and Editors-in-Chief was ``represents one of the explanations, and probably one of the most important one, which interprets the world dominant position of the US in science publication in most of science fields'' \citep[p.319]{braun2005b}.

Another stream of studies focuses on the \emph{gender composition} of editorial boards. Much like the analysis of geographic distribution, these studies aim to examine the gender composition of editorial boards and discern if there are disparities in the representation of men and women within the scientific fields covered by a journal. In such cases, an overrepresentation of one gender among editorial board members can potentially lead to biased paper selection, affecting not only gender balance but also specific subject areas, methodologies, or theories \citep{stegmaier2011, metz2016}. \citet{mauleon2013} suggests that increased participation of women on editorial boards can positively influence the attraction of female researchers to their respective scientific disciplines, because women in gatekeeper positions can be perceived as role models for graduate
students and junior researchers. 
The first analysis of female representation in editorial positions was conducted by \citet{hatfield1995}. They observed the low presence of women in the research sector and questioned whether this pattern extended to the editorial level. Their analysis focused on the gender composition of Editor-in-Chief roles in the 100 most influential clinical medicine journals. They found that in 92 out of 96 journals, the most important editorial positions were occupied by men in 92 out of 96 journals, while only 4 by women. In one case, a woman shared the position with three other men. Subsequent studies on the same topic conducted in different fields have yielded similar findings, indicating male domination in editorial boards and a significant gap between the number of female researchers and their representation on these boards. While there has been an increase in the number of women serving on editorial boards, this change has occurred at a slower rate than the increasing presence of women in scientific fields. Moreover, there are fewer women in editorial boards of the most prestigious journals and in the role of Editor-in-Chief (see \citet{mazov2016} for a review).

More recently, the concept of \emph{Interlocking Editorship (IE)} has emerged as a framework for examining the structural characteristics of editorial board networks. Initially proposed by \citet{baccini2009}, an IE network is defined as a network that arises from the presence of the same individual on the editorial boards of multiple journals. The
underlying idea is that the number of editorial board members shared between two journals can be viewed as an indicator of journal similarity,
i.e., the IE approach measures journal proximity based on common
editorial board membership. Another perspective on the IE network is its utility in identifying scholarly communities, often referred to as "invisible colleges", as well as academic elites. This pertains to editors who hold multiple board positions or occupy central positions within the network, which in turn grants them significant influence over editorial decisions. Notably, \citet{baccini2020} discovered that the interlocking editorship network of journals bears similarities to both the co-citation network and the interlocking authorship network of journals. Consequently, studying journal communities within the IE network yields results akin to those obtained by examining communities in the other two networks.
The IE framework has found application in various research fields through Social Network Analysis (SNA). In addition to the already mentioned economics, these fields include statistics \citep{bacciniet2009}, information and library science \citep{baccini2011, liwei2015, ni2010}, finance \citep{andrikopoulos2015}, knowledge management and intellectual capital fields \citep{teixeira2018}, communication sciences \citep{goyanes2020}, tourism \citep{lockstone2021}. These studies have offered valuable insights into the clustering of journals within specific fields or research areas, as well as the underlying structure of editorial gatekeeping. In recent times, the IE network has also been employed to investigate the geographical distribution of co-editor networks in oncology, revealing a core-periphery geographical structure \citep{csomos2022}. 

\section{Data and research questions}
\label{sec:data}
The main objective of this study is to update and enrich the knowledge
about the composition of editorial boards of economics by studying it on a
database that includes all the 1,516 journals indexed in \textit{EconLit}, with an active editorial board in 2019. \textit{EconLit}, published by the American Economic Association (AEA), provides bibliographic coverage of the major scientific economics-related literature and it is the main source of references in the field of economic
literature worldwide. The list of journals was compiled from AEA website in April 2019 (\url{https://web.archive.org/web/20190716024210/https://www.aeaweb.org/econlit/journal_list.php}). 

The data on the members of the editorial boards
was directly collected from the websites of the journals. For each
member, the following data were manually entered: name and surname,
role, journal name, affiliation if declared. All the information
was manually standardized. For name and surname the manual standardization was conducted after an automatic disambiguation based on string similarities. 

The final database collects data about 60,638 seats,
classified in 477 distinct roles, and occupied by 44,460 scholars. The average number of seats per journal turned out to be 40 and
the average number of seats per scholar, i.e., the mean
rate of participation, was 1,36. 

% The seats associated with an affiliation are are 54,098; 570 seats are held by scholars with multiple affiliations; in these cases, for simplicity, the analysis consider only the first affiliation, i.e. the one listed as first in the journal's website.

The seats associated with an affiliation are 53,964; 1,406 seats are held by scholars with multiple affiliations; in these cases, for simplicity, the analysis consider only the first affiliation, i.e.the one listed as first in the journal's website.

For 6,674 seats held by 6,179 scholars no affiliation was available; they represent respectively 11\% and 13.9\% of the total number of seats and scholars
%6,540 seats held by 6,066 scholars no affiliation was available; they represent respectively 10.8\% and 13.6\% of the total number of seats and scholars. 
The affiliated institutions are 6,081.
%6524 
Each distinct affiliation was associated with a country by using the Google Maps Text Search API and by manually cleaning wrong attributions. In this way, 53,700 affiliations, 96.8\% of total, were associated to 142 different countries. 

The gender was attributed to scholars by using an algorithm, based on the package \pkg{genderize.io},
that considered both the first name and the country of the member's
affiliation, in order to take into account geographical variability in
the association between names and gender (e.g., the name `Andrea' is
mainly attributed to men in Italy but to women in English-speaking
countries). The gender was coded on a binary scale (male -- female) not having
the possibility to obtain self-reported gender data. We apologize to all
those who are represented in the sample and who do not self-identify
along the hetero-normative binary and hope that future studies might have
more resources to contact people individually to report on
self-identified data. We have been able to attribute gender to 39,761 individuals (89.4\% of the total).

The analysis requested the identification of journal editorial leaders , i.e. scholars who have the highest editorial ranking in the journal. Since each journal classifies roles differently, we needed to establish a consistent classification method to identify individuals we will henceforth refer to as `editorial leaders'. 
To this end, two different procedures were adopted. The first one, simply consisted in considering as editorial leaders the scholars classified by journals as Editor-in-Chief, Co-Editor-in-Chief, Deputy Editor-in-Chief or Joint Editor-in-Chief. By this procedure, 981 people are identified as editorial leaders in 687 journals, i.e., the 45.28\% of the 1,516 journals of the database. 
For the rest of journals, a second more complex
procedure was adopted. It consisted of the direct identification of  journal editorial leaders who were classified with a generic name such as Editor, Co-Editor, Director. In these cases the editorial leaders were identified by considering their hierarchical position, first or last, in the list of editorial board members. In a few journals the editorial leadership appear to be held collectively by more than three scholars, grouped in a higher hierarchical position than other members of the boards. Also in these cases we decided to preserve the information and to classify these small groups as editorial leaders.
In sum, a total of 2,893
%2,897 
editorial leaders in 1,448 (95.45\%) journals were identified. Each journal has on average 2 editorial leaders. 

Table \ref{table:1} reports the main quantitative features of the final dataset.

\begin{table}[ht!]
\caption{Editorial boards of economics journals in 2019: data description}
\label{table:1}
\centering
\begin{tabular}{|l|c|}
\hline
\textbf{Elements} & \textbf{n.} \\
\hline
Journals & 1,516 \\
Seats in the editorial boards & 60,638 \\
Distinct scholars & 44,460 \\
Distinct female scholars & 13,282 \\
Distinct affiliations & 6,081 \\
Distinct countries & 142 \\
Distinct roles & 477 \\
Seats without affiliation & 6,674 \\
Seats without country & 8,416 \\
Seats without gender & 5,603 \\
Distinct Editorial Leaders & 2,893 \\
Distinct female Editorial Leaders & 705 \\
Distinct Editorial Leaders seats & 3,010 \\
\hline
\end{tabular}%
\end{table}

As for the research questions, the dataset has been used, firstly, to explore the composition of the editorial boards of economics journals in order to verify their degree of homogeneity in terms of geographic affiliation, institutional affiliation, and gender. This composition is compared with the data on the population of economists, as registered in \citet{repec2023}.  

The second research questions is about social and intellectual similarity among journals. More specifically, we ask whether it is possible to measure the social similarity between pairs of journals and whwther this similarity allows for the identification of clusters of relatively similar journals. To this end, we adopt a network analysis approach, by focusing on relations among journals represented in terms of interlocking editorships (IE). As anticipated in Section \ref{sec:review}, the IE fundamental unit is the scholar holding multiple seats in different journals. The descriptive analysis of the IE permits us to discuss the notions of prestige and editorial power. The social and intellectual similarity among journals is explored by computing a measure of similarity between each pair of journals based on the number of editorial board members shared between them, as in \citet{baccini2020}. 

Three different similarity networks are constructed and compared: the complete IE network, the IE network created by scholars holding at least one editorial leadership position, and the IE network created by scholars who hold many seats but are never editorial leader. This approach allows us to verify if the structure of the similarity networks and clusters of journals inside them are stable when the links among journals are generated by scholars with different editorial power.

The final descriptive research question is about the most central journals and editors within the complete IE network. Finding the most central journals and editors in the IE network helps to identify the most influential editorial gatekeepers. These central nodes often have a significant impact on shaping the field, as they hold editorial power and are connected to a wide range of scholars and journals. This analysis is conducted also in a gender perspective, by exploring also the network generated by women scholars. By identifying these influential gatekeepers, insights can be gained to understand the distribution of influence and potentially address issues related to diversity, inclusion, and concentration of power in academic publishing. 

All these steps allow us to identify who the gatekeepers of economics are and to detect differences related to roles and gender on an unprecedented large scale, providing a comprehensive answer to the fundamental question underlying this research.

The network analysis and visualization were realized with \textsc{Pajek} (version number 5.14) and \textsc{Gephi} (version number 0.9.5). 

\section{Geographic distribution}
\label{sec:geography}

Table \ref{table:2} presents the 10 most represented countries among the affiliations of editorial board members and editorial leaders. The percentage of each country in relation to all editorial roles was calculated from 51,608 seats where country attribution was possible, representing 85.1 percent of the total number of seats. Similarly, for editorial leaders the percentage of each country was calculated using 2,480 seats with attributed countries, representing 82.5\% of the total of editorial leaders seats.
%The percentage of each country in relation to all editorial roles has been calculated on 51,608 seats, which accounts for 85.1\% of the total, where country attribution was possible.

The United States stands out as the most represented country, holding 33.6\% of all seats and 35.4\% of editorial leader seats. It is followed far behind by the United Kingdom with 9.2\% and 9.1\%, respectively. Among the top 10, only countries categorized as influenced or part of Western nations are represented. Moreover, the five most represented countries collectively occupy the majority of seats, accounting for 54.8\% of all editorial seats and 57.9\% of editorial leader seats. While 142 countries have at least one seat in a journal, only 81 countries (approximately 43\%) have at least one editorial leader seat.

\begin{table}[ht!]
\caption{Seats at the editorial tables. The 10 most represented countries.}
\label{table:2}
\centering
\begin{tabular}{|lcc|lcc|}
\hline
\multicolumn{3}{|c|}{\textbf{All Editorial Roles}} & \multicolumn{3}{|c|}{\textbf{Editorial Leaders}}\\
\hline
\textbf{Country} & \textbf{Total} & \textbf{Percentage} & \textbf{Country} & \textbf{Total} & \textbf{Percentage} \\
\hline
United States & 17329 & 33.6 & United States & 879 & 35.4 \\
United Kingdom & 4737 & 9.2 & United Kingdom & 225 & 9.1 \\
Italy & 2154 & 4.2 & Germany & 141 & 5.7 \\
France & 2057 & 4.0 & Italy & 96 & 3.9 \\
Canada & 2007 & 3.9 & Canada & 94 & 3.8 \\
Germany & 1899 & 3.7 & Spain & 72 & 2.9 \\
Spain & 1750 & 3.4 & Australia & 67 & 2.7 \\
Australia & 1653 & 3.2 & France & 66 & 2.7 \\
Turkey & 1254 & 2.4 & Netherlands & 58 & 2.3 \\
Netherlands & 911 & 1.8 & Japan & 48 & 1.9 \\
\hline
\end{tabular}
\end{table}

These results can be compared with data on the actual geographic distribution of economists to determine if there are differences between members of the editorial roles and economists in general. The most readily available data comes from \citet{repec2023}, which collects data about economists in 2023. The three-year difference from our database is short enough to assume that the distribution of economists by country has not changed dramatically in the meantime.

The comparison of the geographic distribution of editorial seats in Table \ref{table:2} with the country affiliation of economists registered in \citet{repec2023} in Table \ref{table:2a} shows that the United States, the United Kingdom, Canada, Australia, Turkey, and the Netherlands are over-represented in the editorial boards of economic journals. All the other countries are instead under-represented. 
%In fact, as it is possible to see in Table \ref{table:2a}, the 20.8\% of economics authors are affiliated to US, the 6.7\% to UK, the 6.5\% to France, the 6.1\% to Germany and 5.8\% to Italy, the 4.3\% to Spain, the 3 \% to Canada, the 2.7\% to Australia, the 2\% to Netherlands, the 1.5\% to Turkey.
In particular, Russia ranks 8th and China 10th among the top 10 most represented countries in RePEc. When examining the geographic distribution of editorial boards, China ranks 12th for all editorial roles with 796 seats and 24th for editorial leaders with 20 seats. Russia, on the other hand, is positioned at 38th with 242 seats for all editorial roles and at 29th with 13 seats for editorial leaders.

\begin{table}[ht!]
\caption{The 16 most represented countries in RePEc (2023).}
\label{table:2a}
\centering
\begin{tabular}{|lcc|lcc|}
\hline
\textbf{Country} & \textbf{Total} & \textbf{Percentage} & \textbf{Country} & \textbf{Total} & \textbf{Percentage} \\
\hline
United States & 11966 & 20.8 & Australia & 1543 & 2.7 \\
United Kingdom & 3820 & 6.7 & China & 1399 & 2.4 \\
France & 3724 & 6.5 & Japan & 1195 & 2.1 \\
Germany & 3520 & 6.1 & Netherlands & 1152 & 2.0 \\
Italy & 3352 & 5.8 & Romania & 1149 & 2.0 \\
Spain & 2450 & 4.3 & India & 1147 & 2.0 \\
Canada & 1692 & 2.9 & Switzerland & 1040 & 1.8 \\
Russia & 1578 & 2.7 & Turkey & 892 & 1.5 \\
\hline
\end{tabular}
\end{table}

As for the country composition of the board of each journal, a concentration metric is developed by calculating the proportion of seats for each country represented on the board. Journals are defined as `highly concentrated in terms of geographic diversity' when a single country holds at least 50\% of the total seats, or of the editorial leader seats. Table \ref{table:3} reports the number and percentage of the highly concentrated journals in terms of geographic diversity.

When considering all editorial roles, 504 journals (33\% of the total) exhibit a high concentration of geographic diversity, that is, they have a significant number of members affiliated with the same country. Of these, 273 journals are associated with the United States, 23 with Turkey, 22 with Spain, 17 with France, 16 with Italy, 14 with Germany, and only 12 with the United Kingdom. These journals can be regarded as nationally based journals, and their presence contributes to the ranking of the most represented countries in the editorial boards, as shown in Table \ref{table:2}. It's worth noting that 12 journals have all their editorial board members affiliated with the same country. Three of them are from the United States  (\emph{American Economist}, \emph{American Law and Economics Review}, \emph{Financial Markets, Institutions and Instruments}).

When considering only editorial leader seats, 967 journals (64\% of the total) are highly concentrated in terms of geographic diversity. Table \ref{table:3} shows that for 860 journals, the editorial leader comes from a single country. The countries with more than 50\% of editorial leaders in each journal taken separately are again the United States (322 journals), the United Kingdom (75), Germany (40), Italy (36), Spain (32), and France (29).

\begin{table}[ht!]
\caption{Seats at the editorial tables. Journals highly concentrated in terms of geographic diversity.}
\label{table:3}
\centering
\begin{tabular}{|c|cc|cc|}
\hline
& \multicolumn{2}{|c|}{\textbf{All Editorial Roles}} & \multicolumn{2}{|c|}{\textbf{Editorial Leaders}}\\
\hline
\textbf{Range} & \textbf{N° Journals} & \textbf{\% Journals} & \textbf{N° Journals} & \textbf{\% Journals} \\
\hline
50\%-59\% & 162 & 11.75 & 9 & 0.59 \\
60\%-69\% & 120 & 8.71 & 47 & 3.10 \\
70\%-79\% & 103 & 7.47 & 36 & 2.37 \\
80\%-89\% & 71 & 5.15 & 12 & 0.79 \\
90\%-99\% & 36 & 2.66 & 3 & 0.20 \\
=100\% & 12 & 0.87 & 860 & 56.73 \\
\hline
\end{tabular}%
\end{table}

The analysis of the geographic distribution of editorial seats in economics journals reveals a predominant presence of scholars affiliated with the United States, both as editorial board members and as editorial leaders. This dominance holds true when considering all journals collectively and when analyzing the boards of individual journals. %This dominance becomes hegemony in the case of the Top Five journals}. 
Furthermore, this presence on editorial boards is disproportionately higher compared to the number of economic authors affiliated with the United States. In contrast, Russia and China are underrepresented on editorial boards relative to the number of economics authors from these countries. These results confirm that, as of 2019, the United States maintains its position as the leading scientific power in economics, with the United Kingdom and other Western countries (particularly Germany, Italy, France, and Spain) following at a significant distance.

\section{Institutional distribution}
\label{sec:institutions}

Shifting our focus to the institutional level, we can determine whether some universities or research centers are more represented than others on editorial boards and assess their degree of concentration. Table \ref{table:5} reports the ten most represented institutions in the editorial boards of economics journals. In this case, the percentage for each institution has been calculated based on 53,964 seats (89\% of the total) for which an affiliation could be attributed. Similarly, the percentage for editorial leader seats is calculated over 2,580 (14.2\%) total editorial leader seats with affiliations.

Table \ref{table:5} reveals that the most represented institution is the University of California, both in all editorial roles and in editorial leader seats. Nevertheless, this result is magnified due to the difficulty in uniformly understanding which campus of the University of California scholars belong to, during the process of standardization of affiliations. Table \ref{table:A} in the Appendix provides affiliations as reported on the websites of journals for a generic `University of California' and its campuses. In any case, the majority of the most represented institutions are located in the United States. There are only two exceptions among all editorial roles: the London School of Economics and the University of Oxford. Among editorial leaders, only the London School of Economics is in the top 10, outside the United States.

In this case, too, affiliation diversity is higher for all editorial roles compared to editorial leaders: there are 6,081 different institutions represented in total seats, whereas editorial leaders are affiliated with only 1,036 institutions (17\%). Moreover, the concentration is slightly lower for all editorial roles compared to editorial leaders: the top 10 institutions collectively represent 8.2\% of total seats for all editorial roles and 11.7\% of editorial leader seats.

\begin{table}[ht!]
\caption{Seats at the editorial tables. The 10 most
represented institutions. (* See the Appendix for data about University of
California).}
\label{table:5}
\centering
\begin{tabular}{|lcc|lcc|}
\hline
\multicolumn{3}{|c|}{\textbf{All Editorial Roles}} & \multicolumn{3}{|c|}{\textbf{Editorial Leaders}}\\
\hline
\textbf{Institution} & \textbf{Total} & \textbf{Percent.} & \textbf{Institution} & \textbf{Total} & \textbf{Percent.} \\
\hline
University of California* & 1091 & 1.32 & University of California* & 46 & 1.78 \\
London School of Economics & 566 & 1.05 & University of Pennsylvania & 29 & 1.12 \\
University of Pennsylvania & 500 & 0.93 & MIT & 29 & 1.12 \\
Harvard University & 478 & 0.89 & University of Chicago & 26 & 1.01 \\
Columbia University & 412 & 0.76 & London School of Economics & 24 & 0.93 \\
New York University & 384 & 0.71 & Harvard University & 24 & 0.93 \\
Michigan State University & 381 & 0.71 & Northwestern University & 22 & 0.85 \\
University of Oxford & 345 & 0.64 & University of Washington & 21 & 0.81 \\
Stanford University & 337 & 0.62 & Stanford University & 21 & 0.81 \\
University of Washington & 329 & 0.61 & Yale University & 20 & 0.78 \\
& & &	New York University & 20 & 0.78 \\
& & & Columbia University & 20 & 0.78 \\
\hline
\end{tabular}
\end{table}

The lower concentration in all editorial roles compared to editorial leader seats is confirmed when the analysis is conducted at the individual journal level. In this case, the concentration of institutions in each journal is calculated by considering the proportion of seats from each institution on the board over the total number of seats on each journal's board. A journal is considered `highly concentrated in terms of institutional diversity' if at least 50\% of its editorial board seats are occupied by members from a single institution. Table \ref{table:6} reports the number and percentage of highly concentrated journals in terms of institutional diversity. Out of 1,516 journals, only 30 (2\%) exhibit a high concentration of members affiliated with the same institution. None of these institutions are among the Top 10 most represented institutions in Table \ref{table:5}. Only three journals have all editorial members coming from the same institution: \textit{Economic Outlook} from Curtin University (Australia), \emph{Journal of Islamic Economics, Banking and Finance} from the University of Bahrain (Bahrain), and \emph{Strategic Finance} from Corvinus University of Budapest (Hungary). 

Focusing on editorial leader seats, there are 801 high-concentration journals (52.8\% of the total), among which 781 journals exhibit an extreme concentration with editorial leaders belonging to a single institution. The three most represented institutions among the journals with extreme concentration are the University of California (15 journals), the University of Bologna (7), and Florida State University (7).

\begin{table}[ht!]
\caption{Seats at the editorial tables. Journals high concentrated in terms of institutional diversity.}
\label{table:6}
\centering
\begin{tabular}{|c|cc|cc|}
\hline
& \multicolumn{2}{|c|}{\textbf{All Editorial Roles}} & \multicolumn{2}{|c|}{\textbf{Editorial Leaders}}\\
\hline
\textbf{Range} & \textbf{N° Journals} & \textbf{Percentage} & \textbf{N° Journals} & \textbf{Percentage} \\
\hline
50\%-59\% & 12 & 0.79 & 4 & 0.26 \\
60\%-69\% & 6 & 0.40 & 10 & 0.66 \\
70\%-79\% & 3 & 0.20 & 6 & 0.40 \\
80\%-89\% & 2 & 0.13 & 0 & 0.00 \\
90\%-99\% & 4 & 0.26 & 0 & 0.00\\
=100\% & 3 & 0.20 & 781 & 51.52 \\
\hline
\end{tabular}%
\end{table}

Therefore, it is possible to state that it is difficult to identify `hegemony' by any particular institution in the editorial boards of economics journals. Instead, there is a widespread representation of US universities and some UK universities.

\section{Gender composition}

The analysis of gender composition in economics journal editorial boards requires a brief contextualization. In general, women are underrepresented in the field of economics, with a more significant disparity at higher academic positions. As documented by \citet{lundberg2019}, the field of economics became substantially less male-dominated during the 1980s and 1990s, but this growth in female representation has stalled. The proportion of female assistant professors and PhD students has remained relatively constant since the mid-2000s, and is around 25\%. In contrast, women's representation at senior levels has been increasing but remains at nearly 14\% as of 2017. This difference between the initial positions and higher positions held by women within the profession is consequently referred to as a kind of ‘glass ceiling'. 
The database used here enables us to explore whether this vertical gender segregation is reflected in the composition of editorial boards.

Gender information could be assigned to scholars occupying 55,035 seats, representing 90.76\% of the total available seats. For editorial leaders, 2,781 were gender-identified, accounting for 92.5\%. As shown in Table \ref{table:8}, women occupy approximately 25\% of the total available seats and editorial leader seats, while men account for the remaining 75\%. At first glance, it cannot be said that there is vertical segregation: women are underrepresented within editorial boards or among editorial leaders only because there are fewer of them within the profession.

\begin{table}[ht!]
\caption{Seats at the editorial tables. Gender composition.}
\label{table:8}
\centering
\begin{tabular}{|l|cc|cc|}
\hline
& \multicolumn{2}{|c|}{\textbf{All Editorial Roles}} & \multicolumn{2}{|c|}{\textbf{Editorial Leaders}}\\
\hline
\textbf{Gender} & \textbf{Total} & \textbf{Percentage} & \textbf{Total} & \textbf{Percentage} \\
\hline
Female & 13282 & 24.13 & 705 & 25.35 \\
Male & 41753 & 75.87 & 2076 & 74.65 \\
\hline
\end{tabular}
\end{table}

However, there could be ‘horizontal segregation' with women being more represented in certain fields or journals. To investigate this hypothesis, the gender composition of editorial boards for each journal is analyzed separately. A journal can be considered to have a ‘high male composition of seats' if more than 50\% of its seats are held by men. The number and percentage of journals with a high male composition of editorial boards is reported in Table \ref{table:9}: 1,322 journals, i.e.  87\% of economics journals, have a high male composition of board. Furthermore, if the threshold is set at 75\% male, 725 journals (47.8\%) exceed this threshold. Table \ref{table:10} presents the top 10 journals with the highest proportion of women on their editorial boards, of which three journals are focused on gender or feminist topics. 

Regarding editorial leader seats, there are 974 journals (64.2\% of economics journals) where at least 50\% of editorial leader seats are held by men. Among these, 873 journals (57.6\%) have more than 75\% of editorial leader seats occupied by men, and 836 journals (55\%) have men holding all 100\% of editorial leader seats. In contrast, only 262 journals (17\%) have more than 50\% of editorial leader seats occupied by women, and among these, 235 journals (15.5\%) have 100\% of their editorial leader seats held by women.

\begin{table}[ht!]
\caption{Seats at the editorial tables. Journals with high male composition of seats.}
\label{table:9}
\centering
\begin{tabular}{|c|cc|cc|}
\hline
& \multicolumn{2}{|c|}{\textbf{All Editorial Roles}} & \multicolumn{2}{|c|}{\textbf{Editorial Leaders}}\\
\hline
\textbf{Range} & \textbf{N° Journals} & \textbf{\% Journals} & \textbf{N° Journals} & \textbf{\% Journals} \\
\hline
50\%-59\% & 122 & 8.05 & 15 & 0.99 \\
60\%-69\% & 268 & 17.68 & 51 & 3.37 \\
70\%-79\% & 426 & 28.10 & 54 & 3.56 \\
80\%-89\% & 357 & 23.55 & 15 & 0.99 \\
90\%-99\% & 114 & 7.52 & 3 & 0.20 \\
= 100\% & 35 & 2.31 & 836 & 55.15 \\
\hline
\end{tabular}%
\end{table}

\begin{table}[ht!]
\caption{Seats at the editorial tables. The 10 journals with more female presence.}
\label{table:10}
\centering
\begin{tabular}{|lcc|}
\hline
\multicolumn{3}{|c|}{\textbf{All Editorial Roles}}\\
\hline
\textbf{Journal Name} & \textbf{\% Women} & \textbf{Editorial Seats} \\
\hline
Feminist Economics & 81.91 & 94 \\
Monetary Policy and the Economy & 77.78 & 18 \\
Indian Journal of Gender Studies & 76.00 & 25 \\
Indiana Business Review	& 75.00 & 4 \\
Journal of Economic Perspectives & 68.75 & 16 \\
International Business and Global Economy & 66.67 & 24 \\
Pennsylvania Economic Review & 66.67 & 6 \\
Journal of Economic Literature & 64.71 & 34 \\
Studies in Family Planning & 64.29 & 28 \\
Focus on European Economic Integration & 62.50 & 16 \\
\hline
\end{tabular}
\end{table}

Summing up, the overall presence of women in the editorial boards of economics journals, whether for all editorial boards or for editorial leaders, is similar to that observed in academic positions, accounting for approximately 25\% of total seats in both cases. However, women are more prominent in some journals than in others: in 47\% of journal boards and 57.6\% of editorial leader seats, the presence of women falls below 25\%. These results suggest that, unlike academic positions, there is not a form of ‘vertical segregation' or ‘glass ceiling' but ‘horizontal segregation' of women on some editorial boards.  In this case, the horizontal segregation is probably related to the fact that some topics are more women-intensive, or that only a few journals care about gender balance in selecting their editorial board members. This last hypothesis will be checked in the following Sections.

\section{Interlocking Editorship Networks}
\label{prestige}

Up to this point, the unit of analysis has been the editorial board seat. However, a scholar can hold multiple seats simultaneously. In fact, the composition of an editorial board contributes to a journal's prestige. Therefore, journals appoint `famous' or `influential' scholars to enhance their reputation and attract the 'best' \citep{baccini2010}. Conversely, editors of journals with strong reputations wield significant power \citep{faria2005}. Hence, scholars tend to accept multiple roles on different editorial boards

\begin{figure}[ht]
\centering
\includegraphics[scale=0.8]{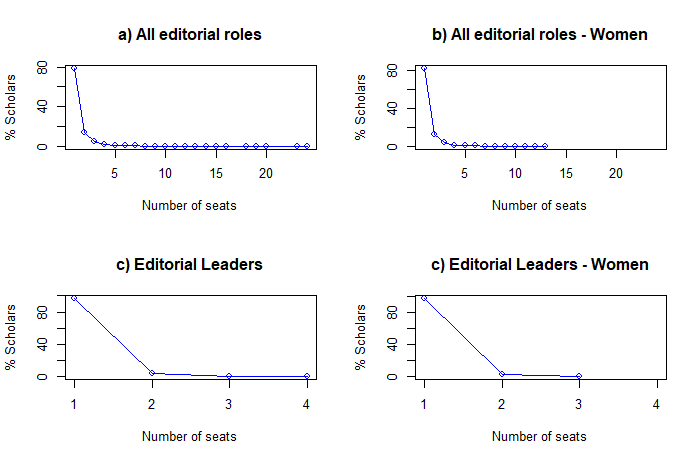}
\caption{Distribution of scholars according to the number of seats held in the editorial boards of economics journals.}
\label{figure:1}
\end{figure}

The descriptive analysis of the distribution of scholars based on the number of seats they hold is presented in Figure \ref{figure:1}. When considering all editorial roles, approximately 79\% of scholars occupy only one seat. The remaining 9,520 scholars (21.4\%) hold more than one seat, with a maximum of 24 seats held by a single scholar. In the case of editorial leaders, however, only 110 scholars (3.7\%) hold more than one seat, with a maximum of 4 seats held by the same person. Moreover, women tend to hold fewer seats simultaneously. For all editorial roles, only 17.6\% (1,837 out of 10,424) of women hold more than one seat, with a maximum of 13 seats held by a single person. In the case of editorial leaders, the distribution is quite similar between genders, with a similar 3.5\% of women scholars (24 out of 681) holding more than one editorial leader seat, with a maximum of 3 editorial leader seats held by the same person.

In summary, there are `prominent' or `prestigious' economists who sit on many editorial boards, but very few scholars act as editorial leaders in multiple journals. This suggests that editorial leaders hold the real editorial power and have a high editorial workload, making it difficult for a scholar to serve as an editorial leader in more than one journal. On the other hand, the role of a member of an editorial board may appear honorary for scholars -- predominantly men -- sitting on the boards of many journals. In turn, it may seem that being selected as an editorial board member can enhance a journal's prestige rather than simply conferring effective power to the scholar on the board.

%Table \ref{table:13} reports the top-10 of multiple editors, i.e., the 10 scholars holding  the highest number of seats in the editorial boards of economics journals. On the left side of the table, all editorial roles are considered, while the right side shows the scholars who hold more seats, including at least one as editorial leader. 

%\begin{table}[ht!]
%\caption{Seats at the editorial tables. The 10 scholars holding the highest number of seats.}
%\label{table:13}
%\centering
%\begin{tabular}{|lc|lc|}
%\hline
%\multicolumn{2}{|c|}{\textbf{All Editorial Roles}} & \multicolumn{2}{|c|}{\textbf{At Least One Editorial Leader}}\\
%\hline
%\textbf{Name and Surname} & \textbf{N° of seats} & \textbf{Name and
%Surname} & \textbf{N° of seats} \\
%\hline
%Amartya K. Sen & 24 & Mohammad Kabir Hassan & 20 \\
%Barry Eichengreen & 23 & Andrés Rodríguez Pose & 19 \\
%Mohammad Kabir Hassan & 20 & Douglas J. Cumming & 18 \\
%Andrés Rodríguez Pose & 19 & Wing-Keung Wong & 16 \\
%Douglas J. Cumming & 18 & Iftekhar Hasan & 15 \\
%Wing-Keung Wong  & 16 & Giovanni Dosi & 14 \\
%Peter Nijkamp & 16 & Dani Rodrik & 14 \\
%Iftekhar Hasan & 15 & Brian M. Lucey & 14 \\
%Geoffrey M. Hodgson & 15 & Bruno S. Frey & 13 \\
%Giovanni Dosi & 14 & James J. Heckman& 13 \\
%\hline
%\end{tabular}
%\end{table}

\subsection{Social and intellectual similarities in the interlocking editorship network of journals}
\label{IE}

The mentioned difference among `powerful' editorial leaders and `prestigious scholars' who sit on many boards suggests analyzing whether there are different structural characteristics in the networks they create. 
In general, an interlocking editorship approach can be used to explore the different structural properties of the networks generated by the crossed presence of scholars in many boards contemporaneously. An interlocking editorship network is a bipartite network with two sets of nodes, editorial board members and journals, and edges linking members to the journals where they sit. 
Specifically, three interlocking editorship networks are constructed. The first one (`Complete network') is the standard interlocking editorship network, created by considering all editors on journal boards, regardless of their role. The second is the interlocking editorship network formed by scholars who do not hold an editorial leader position (`No-EL network'). The third interlocking editorship network is created by the subset of scholars who hold at least one editorial leader position (`EL network').

\begin{sidewaysfigure}
\centering
\vline
\subfloat [Complete network]{\includegraphics[scale=0.21]{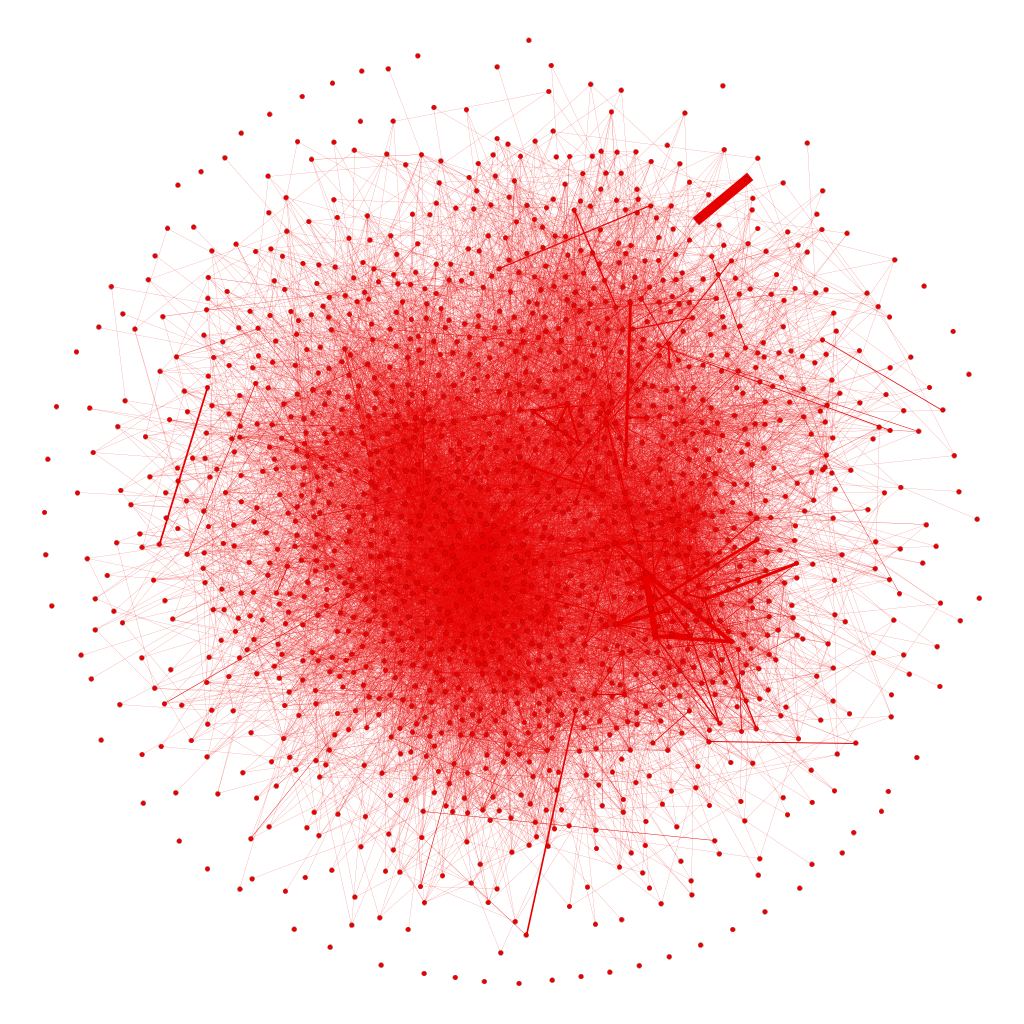}}
\vline
\subfloat [No-EL Network]{\includegraphics[scale=0.21]{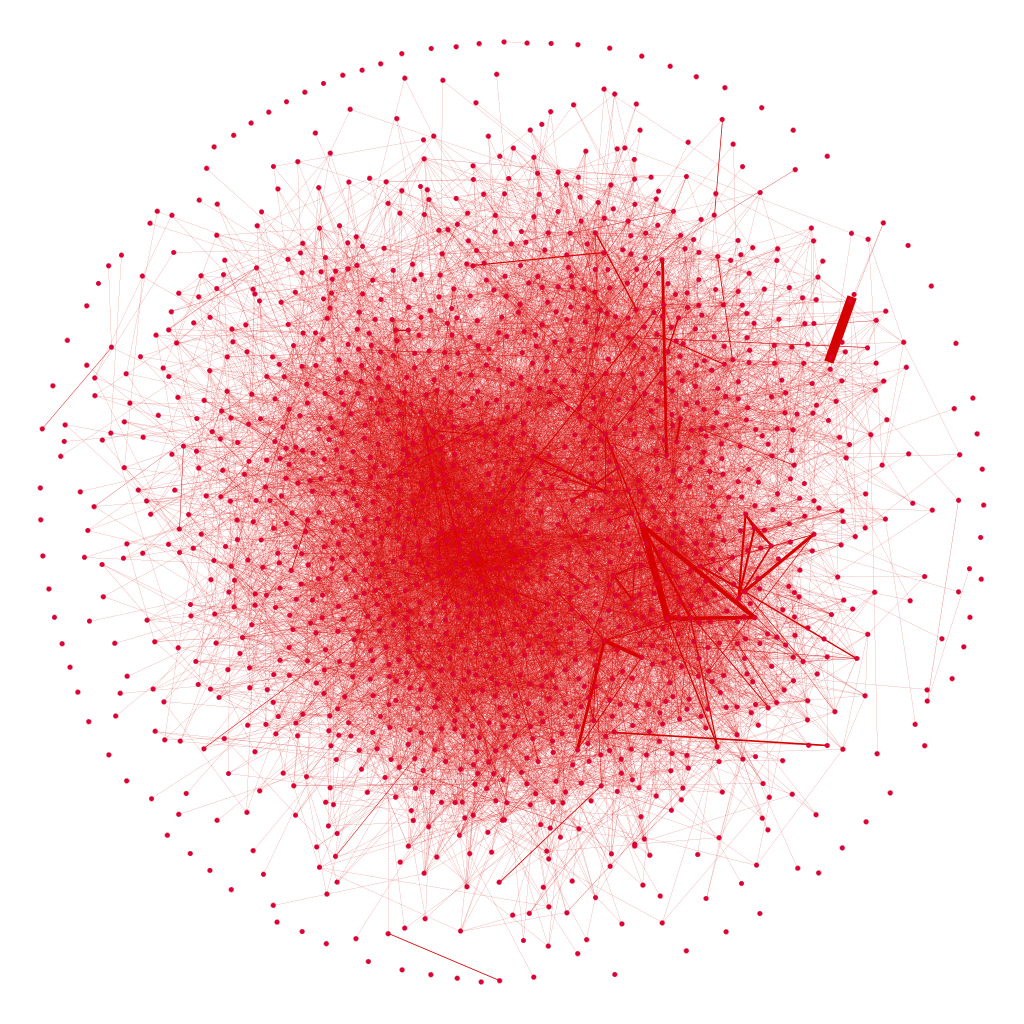}}
\vline
\subfloat [EL network]{\includegraphics[scale=0.21]{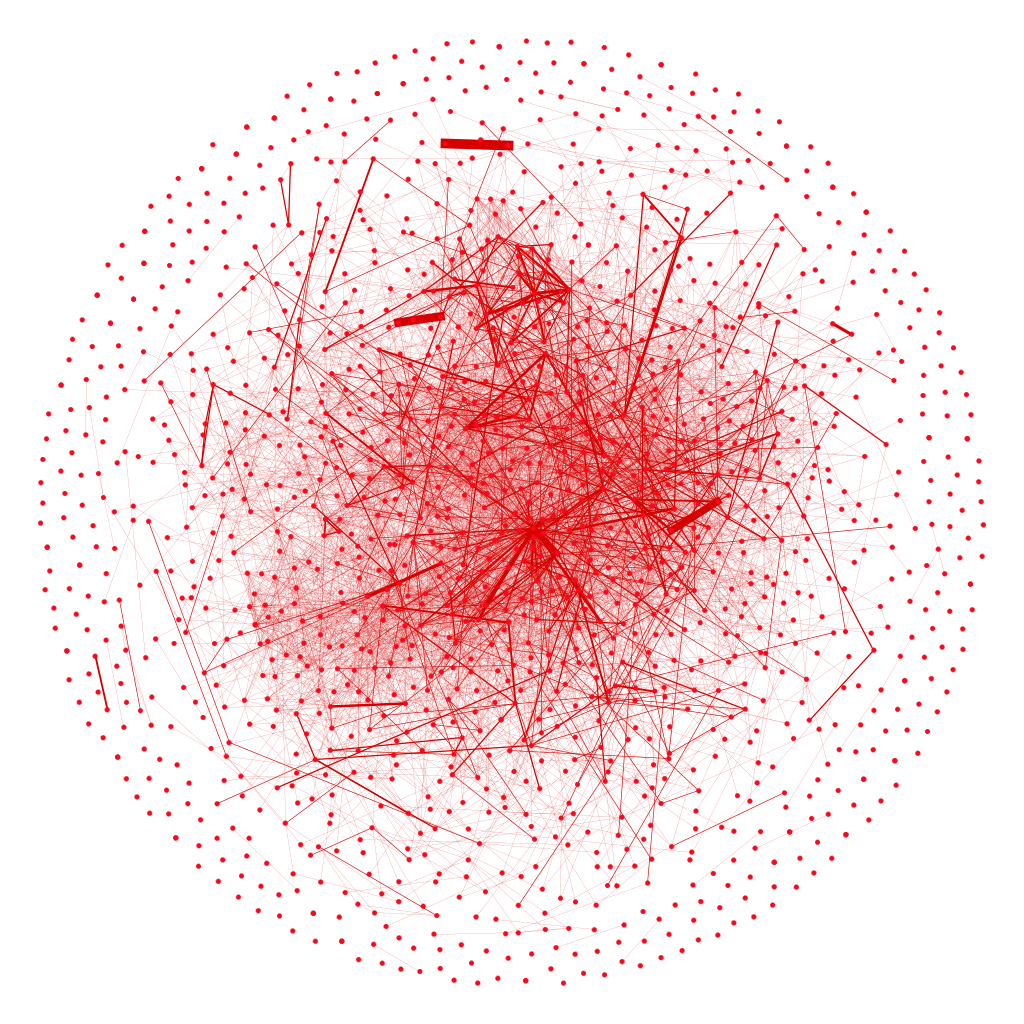}}
\vline
\caption{The interlocking editorship networks. A link between two nodes indicates that two journals share at least one member of editorial boards. The size of the edges is proportional to the number of common scholars. The Complete network is built by considering all the editorial board members; the No-EL network is generated by the subset of scholars who do not hold an editorial leader position; the EL Network is generated by scholars who hold at least one editorial leader position.}
\label{figure:2}
\end{sidewaysfigure}

The exploratory analysis focuses on the projected one-mode network of journals of the three IE networks: two nodes, representing journals, are connected by an edge if they share at least one scholar in their editorial board. The weight of the edge is represented by the number of common scholars. These three networks are represented in Figure \ref{figure:2} by using the Fruchterman-Reingold algorithm on \textsc{Gephi} \citep{bastian2009}. 

\begin{table}[ht!]
\caption{Basic statistics of the IE networks of journals.}
\label{table:14}
\centering
\begin{tabular}{|l|c|c|c|}
\hline
& \textbf{Complete Network} & \textbf{No-EL network} & \textbf{EL Network} \\
\hline
N. of journals & 1,516 & 1,516 & 1,516 \\
Number of links between journals & 20,321 & 15,995 & 6,149 \\
Lowest value of line & 1 & 1 & 1 \\
Highest value of line & 173 & 169 & 17 \\
Number of links with value =1 & \makecell{15,904 \\ (78.26\%)} & \makecell{12,801 \\ (80.03\%)} & \makecell{5,444 \\ (88.53\%)} \\
Number of links with value =2 & \makecell{2,655 \\ (13.06\%)} & \makecell{1,966 \\ (12.29\%)} & \makecell{504 \\ (8.19\%)} \\
Number of links with value \textgreater2 & \makecell{1,762 \\ (8.68\%)} & \makecell{1,228 \\(7.68\%)} & \makecell{201 \\ (3.23\%)} \\
Density & 0.017 & 0.013 & 0.005 \\
Average Degree & 26.81 & 21.10 & 8.11 \\
Betweenness Centralization & 0.043 & 0.043 & 0.051 \\
Number of weak components & 50 & 74 & 372 \\
N. of journals in the largest component & \makecell{1,467 \\ (96.77\%)} & \makecell{1,442 \\ (95.11\%)} & \makecell{1,103 \\ (72.75\%)} \\
Isolated journals & \makecell{46 \\ (3.03\%)} & \makecell{69 \\ (4.55\%)} & \makecell{314 \\ (20.71\%)} \\
\hline
\end{tabular}
\end{table}

Table \ref{table:14} shows that the Complete network is the most connected and dense, while the EL network is the least connected and dense. Moreover, the distribution of the link weights, i.e. the distribution of number of common editors between pairs of journals, indicates the  highest values for the Complete network, intermediate for the No-EL network, and the lowest for the EL network. The proportion of total links with a weight of $1$, indicating that a pair of journals shares only one board member, is 78.26\% in the Complete network, 80.03\% in the No-EL network, and 88.53\% in the EL network. The highest value of a link is respectively 173, 169, and 17 in the three networks. The values of network betweenness centralization are quite similar but slightly higher in the third case, suggesting that there are more central actors in the EL network.

The Complete network and the No-EL network are less fragmented than the EL network. The EL network is composed of 372 weak components compared to 74 in the No-EL network and 50 in the Complete network. Moreover, the largest component of the EL network is smaller than in the other two networks; it contains 72.75\% of the journals compared to 95.11\% in the No-EL network and 96.77\% in the Complete network. Finally, in the Complete network, isolated journals, i.e., journals without any common board member with other journals, are only 46 (3.03\%), whereas there are 69 (4.55\%) journals in the No-EL network and 314 (20\%) in the EL network.

The distinct characteristics of the three networks analyzed so far suggest that editorial leaders have varying levels of involvement in the formation of interlocking editorship networks. Editorial leaders likely hold greater editorial power, have a higher workload, making it more difficult to serve on multiple editorial boards. This probably explains the lower density and greater fragmentation of the EL network.
However, it cannot be excluded that editorial leaders exercise their editorial power also by guiding the selection of the other members of editorial boards.

These different structural properties of the three IE networks suggest to explore whether the three networks are also globally different and if they are composed by different communities of journals. More precisely, the three IE have different underlying social structures formed respectively by all the editorial boards members, by the no-EL members and by EL. The question is whether these social structure are globally different or not, and if they formed or not different clusters of journals.

To this end, Jaccard similarities between pairs of journals are computed as in \citet{baccini2020} for the three networks. 
More specifically, if $A_i$ and $A_j$ represent the sets of board members of the $i$-th and $j$-th journal, the Jaccard coefficient is defined as
\begin{equation}
J_{ij}=\frac{| A_i \cap A_j |}{| A_i \cup A_j |}\ ,
\end{equation}
where $\mid\cdot\mid$ denotes the cardinality of a set. It is apparent that $0\le J_{ij}\le 1$. Hence, the similarity between two journals is proportional to the number of board members they share: when two journals have exactly the same set of editors, i.e.\ when $A_i=A_j$, the maximum similarity $J_{ij}=1$ occurs. In contrast, the minimum similarity $J_{ij}=0$ is achieved when two journals have no common editors, i.e.\ when $A_i\cap A_j=\emptyset$. 
In the complete network, similarities are computed by considering all the editorial board members. In the other two networks, Jaccard similarities are computed by considering the appropriate sets of editors: the No-EL and EL respectively.

Similarity among journals are organized in three different similarity networks. In view of conjecturing about the global difference among the networks, it is possible to compute the generalized distances correlations suggested by \citet{szekely2007}. It is defined in the interval $[0,1]$. Values close to zero indicate no or very weak association between a pair of network; larger values indicate a stronger association. The distance correlations were evaluated in the \proglang{R}-computing environment \citep{R_soft} by using the \code{dcor} functions of the package \pkg{energy}. 

Table \ref{table:17} reports the generalized distance correlations between the three similarity matrices of the three networks. The generalized distance correlations allow to test whether the information obtained changed when networks are built by considering the links generated by different sets of scholars. The very high values of distance correlation, higher than $0.9$, indicate that the distance between the three networks is very low. In other words: the three networks obtained by using different sets of scholars have very similar structure and convey the same information about the connections among economics journals.

\begin{table}[ht!]
\caption{Generalized distance correlation between networks of journals.}
\label{table:17}
\centering
\begin{tabular}{|l|c|c|c|}
\hline
& \textbf{Complete network} & \textbf{No-EL network} & \textbf{EL network} \\
\hline
\textbf{Complete network} & 1 & 0.999 & 0.936 \\
\textbf{No-EL network} & & 1 & 0.926 \\
\textbf{EL network} & & & 1 \\
\hline
\end{tabular}
\end{table}

In fact, the Complete network incorporates the structure of both the No-EL and EL networks. Therefore, it is possible to measure the contributions of these two networks to the complete one, by using the partial distance correlation proposed by \citet{székely2014}. It measures the degree of association between the similarity matrix of the complete network and one of the two other network, by removing the effect of the other. The partial distance correlations were evaluated in the \proglang{R}-computing environment \citep{R_soft} by using the \code{pdcor} of the package \pkg{energy}. The computed partial distance correlation between the complete network and the No-EL network, by removing the effect of EL network, is $0.987$; while the partial distance correlation between the complete network and the EL network, by removing the effect of No-EL network, is $0.776$. Thus, it can be concluded that the contribution of the No-EL network to the complete network is greater than that of the EL network. 

To corroborate these results, a comparative analysis of the communities or clusters  surrounding the three networks has been conducted. Communities are searched by using the Louvain algorithm \citep{Louvain} and the Leiden algorithm \citep{Traag} based on modularity, both available in the in the package \pkg{igraph} of the \proglang{R}-computing environment \citep{R_soft}. Table \ref{table:15} reports the number of clusters detected with the two algorithms, the values of modularity and quality. Modularity and quality measure how effectively a network is partitioned into distinct communities, by comparing the relative density of edges inside communities with respect to edges between distinct communities. The range of modularity and of quality is $[-1,1]$. A value of $-1$ indicates that there are no edges connecting nodes within communities, whereas a value of $1$ indicates that all edges of the network are within communities and no edges exist between communities. The number of communities detected by using one or the other algorithm is nearly identical. More precisely, for measuring the association between the communities detected through the two algorithms, the values of Rand index \citep{Rand} are computed and reported in the last column of Table \ref{table:15}. These values are very near to the maximum value of $1$ and indicate that both algorithm generate nearly identical results.

\begin{table}[ht!]
\caption{Communities and modularity values in the interlocking editorship networks of economics journals.}
\label{table:15}
\centering
\begin{tabular}{|l|c|c|c|c|c|}
\hline
\textbf & \multicolumn{2}{c|}{\textbf{Louvain Algorithm}} & \multicolumn{2}{c|}{\textbf{Leiden algorithm}} & \textbf{Rand}\\
\textbf{Network} & \textbf{n. of Clusters} & \textbf{Modularity} & \textbf{n. of Clusters} & \textbf{Quality} & \textbf{Index}\\
\hline
Complete network & 65 & 0.51 & 68& 0.52 & 0.94\\
No-EL network  & 89 & 0.52 & 92 & 0.53& 0.94 \\
EL network & 394 & 0.68 & 392 & 0.68 & 0.96 \\
\hline
\end{tabular}
\end{table}

As expected, the number of communities is much higher in the less dense EL network compared to the other two networks. The EL network has also a higher modularity/quality value than the complete and No-EL network: this notwithstanding the connections between nodes within communities are denser than connections between nodes of different communities in all the three networks. These results indicate that the three networks can be partitioned in clusters of journals that have a relatively high degree of social similarity \citep{baccini2020}.

The problem is now to verify if the communities of journals detected in the three networks are associated. The communities obtained in the three networks with the two different algorithms are compared by using again the Rand index, whose values are reported in Table \ref{table:16}. These values indicate a strong association among the communities obtained in the three networks.

\begin{table}[ht!]
\caption{Values of Rand index for the association between communities detected in the networks of economics journals by using Louvain and Leiden algorithms.}

\label{table:16}
\centering
\begin{tabular}{|l|c|c|c|}
\hline
\textbf & \multicolumn{3}{c|}{\textbf{\textit{Louvain Algorithm}}} \\
& \textbf{Complete network} & \textbf{No-EL network} & \textbf{EL network} \\
\hline
\textbf{Complete network} & 1 & 0.91 & 0.89 \\
\textbf{Network of No EL} & & 1 & 0.89 \\
\textbf{Network of EL} & & & 1 \\
\hline
\textbf & \multicolumn{3}{c|}{\textbf{\textit{Leiden Algorithm}}} \\
& \textbf{Complete network} & \textbf{No-EL network} & \textbf{EL network} \\
\hline
\textbf{Complete network} & 1 & 0.93 & 0.88 \\
\textbf{Network of No EL} & & 1 & 0.89 \\
\textbf{Network of EL} & & & 1 \\
\hline
\end{tabular}
\end{table}

In sum, the comparison of the three networks of journals reveals that the EL Network is more fragmented that the others, but all the three networks exhibit highly correlated structures. Furthermore, the three networks can be partitioned in communities that are also highly correlated. These results suggest that members of the editorial boards generate similar connections among journals regardless of their role, but with varying degrees of intensity. Scholars who hold the position of editorial leaders, probably due to the workload requested by their position, tend to be involved in fewer journals and therefore contribute less to the connections among journals. But the connections they generated are structurally similar to the most numerous connections generated by the other members of the boards. 

The association between network structures and communities can be explained by the presence of a significant degree of social homophily within each board: indeed when members with different roles are considered, they tend to generate similar connections among journals, resulting in similar communities. This indicates that links between pairs of journals tend to be redundant, generated both by editorial leaders and by other members of the board.

Although communities of journals are properly defined in terms of social similarity, it can be suggested that social similarity goes hand in hand with intellectual similarity, as documented by \cite{baccini2020} for economics journals in 2006.
It can be conjectured that editorial leaders of economics journals have an indirect role in defining the network structures: they exercise their power by selecting as editorial board members scholars socially and intellectually `similar' to them, who have time to be part of many boards by reinforcing or creating new links with other journals. As documented in previous studies \citep{baccini2009,baccini2010,baccini2020} the different communities detected in the interlocking editorship network gather not only different fields of economics, but also groups of highly specialized journals or groups of journals sharing a common methodological approach to economics. To give interpretive substance to the notion of social and intellectual homophily requires a fine-grained analysis of the main characteristics of the communities identified, which is beyond the scope of this paper.

\subsection{The most central journals in the interlocking editorship network}
\label{journals}

The Complete network can be used to highlight the most central journals, which, as mentioned earlier, is vital for potentially identifying the most influential gatekeepers. To this end, three standard measures of centrality of nodes are computed: degree, betweenness, and closeness centrality. These three measures generated three rankings of journals highly correlated: the highest correlation is between the All degree rank and both closeness rank (0.92), intermediate value is between degree rank and betweenness rank (0.79), while the lowest correlation is between betweenness rank and closeness rank (0.77). Therefore, for the sake of simplicity, the discussion will focus on journal degrees.

\begin{figure}[ht!]
\centering
\includegraphics[scale=0.7]{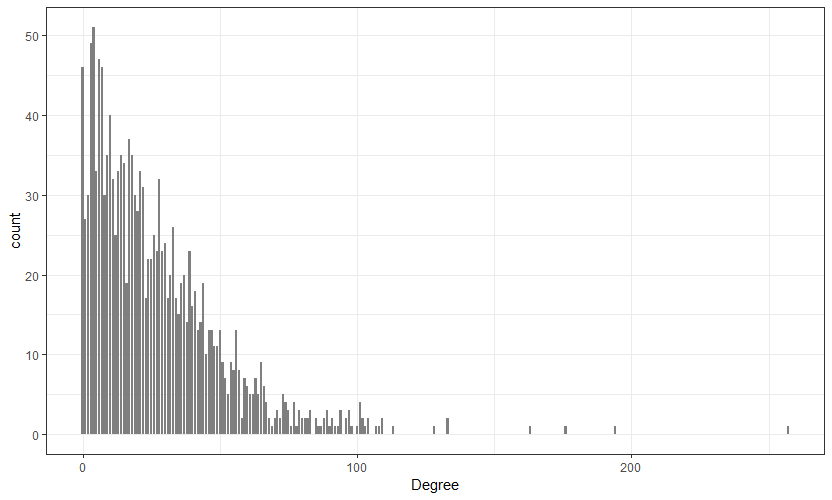}
\caption{Distribution of economics journals according to their degree in the interlocking editorship network.}
\label{figure:3}
\end{figure}

Figure \ref{figure:3} illustrates the degree distribution of the journals, where the degree of a journal represents the number of journals linked to it by at least one common editor. Figure \ref{figure:3} reveals that the distribution is right-skewed, with 15 journals having a degree greater than 100. The median degree is approximately 4, the average degree is 26.8.

Table \ref{table:18} presents the top 10 journals with the highest degree. \emph{Economics: The Open-Access, Open-Assessment E-Journal} id in the first position, linked to 257 other journals by at least one common editor, followed by the \emph{Journal of Risk and Financial Management}, connected to 194 other journals. 

All these journals are indexed in Scopus or Web Of Science; three (\textit{Emerging Markets Finance and Trade}, \textit{Journal of International Business Studies}, \textit{Management Science}) are also in preminent position in Scopus ranking of Journals. It can be only conjectured that the composition of editorial boards of these journals may be guided by a `strategy' aimed at gaining prestige through the selection of `prestigious' members, i.e. members already sitting in other boards.

\begin{table}[ht!]
\caption{Economics journals with the highest degree.}
\label{table:18}
\centering
\begin{tabular}{|l|cc|}
\hline
\textbf{Journal name} & \textbf{Degree} & \textbf{Rank Degree} \\
\hline
Economics: The Open-Access, Open-Assessment E-Journal &	257 & 1 \\
Journal of Risk and Financial Management & 194 & 2 \\
Panoeconomicus & 176 & 3 \\
Emerging Markets Finance and Trade & 163 & 4 \\
International Economics and Economic Policy & 133 & 5 \\
Management Science & 133 & 5 \\
Pacific Economic Review & 128 & 7 \\
Journal of International Business Studies & 113 & 8 \\
Review of International Economics & 109 & 9  \\
Structural Change and Economic Dynamics & 109 & 9  \\
\hline
\end{tabular}
\end{table}

\begin{figure}[ht]
\centering
\includegraphics[scale=0.7]{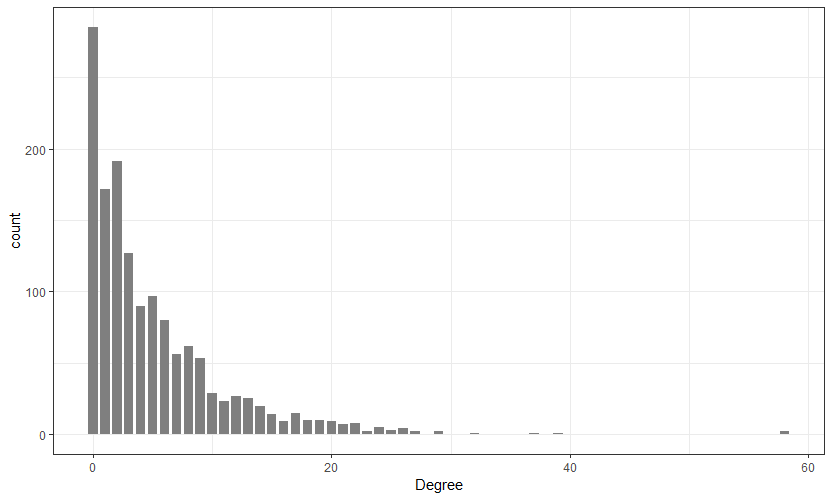}
\caption{Distribution of economics journals according to their degree in the interlocking editorship network built by considering only female scholars.}
\label{figure:4}
\end{figure}

A different scenario emerges when examining the network of journals generated solely by female scholar. Recall that, as seen in Section 6, women represents 25\% of the total number of seats available and they hold fewer seats simultaneously. In this case as well, the rankings of journals based on various centrality measures exhibit high correlations (all exceeding 0.8). The degree distribution remains right-skewed, but the maximum degree is considerably lower, as depicted in Figure \ref{figure:4}. The average degree drops to 5, indicating that, on average, one journal is linked to five other journals; 285 journals (20\%) are isolated.

Table \ref{table:20} presents the most central journals in the interlocking editorship network generated by women. Only three out of ten journals are also among the most central in the complete network: \emph{Management Science}, \emph{Journal of International Business Studies} and \emph{Panoeconomicus}. The remaining seven journals specialize in specific economic perspectives or topics, such as feminist economics, business, human development, or behavioral economics. Some are published by scholarly societies that prioritize gender diversity representation, including the \textit{Journal of Economic Literature} by the American Economic Association and the \textit{Italian Economic Journal}, which is the journal of the Italian Economic Association.

\begin{table}[ht!]
\caption{Economics journals with highest degree in the interlocking editorship network built by considering only female scholars.}
\label{table:20}
\centering
\begin{tabular}{|l|cc|}
\hline
\textbf{Journal name} & \textbf{Degree} & \textbf{Rank Degree} \\
\hline
Management Science & 58 & 1 \\
Feminist Economics & 58 & 1 \\
Journal of International Business Studies & 39 & 3 \\
Journal of Business Research & 37 & 4 \\
Journal of Economic Literature & 32 & 5 \\
Management and Organization Review & 29 & 6 \\
Journal of Human Development and Capabilities & 29 & 6 \\
Italian Economic Journal & 27 & 8 \\
Review of Behavioral Economics & 27 & 8 \\
Panoeconomicus & 26 & 10 \\
\hline
\end{tabular}
\end{table}

These findings validate the previous observation that women tend to hold fewer seats simultaneously because economics ‘prestigious scholars', selected for editorial boards, are predominantly men. It further underscores the existence of a form of ‘horizontal segregation' among women, resulting in their concentration within specific journals.

\section{Editorial leaders: prestige or editorial power?}
\label{EIC}

So far, it has been observed that journals are highly connected through editors who serve on numerous boards. To understand who the scholars that create these connections are, it is possible to use the network of editors. In this network, nodes represent members of editorial boards and the weight of each edge indicates the number of editorial boards on which the pair of scholars sit together. The presence of the same person on the editorial board of more than one journal can be analysed to study the `editorial power' and `academic prestige' of scholars. For prioritizing editorial power over academic prestige, the network is built by considering scholars who serve as editorial leader of at least one journal. The most central scholar in this network probably hold the most editorial power.

\begin{figure}[ht]
\centering
\includegraphics[scale=0.7]{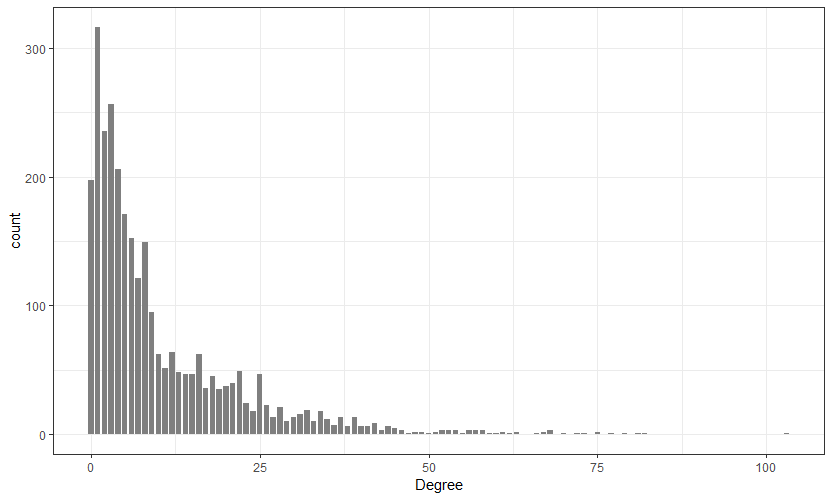}
\caption{Degree distribution of scholars who serve as editorial leader in at least one economics journal. ``Degree" refers to the number of links to other scholars who also serve as editorial leader in at least one economics journal.}
\label{figure:5}
\end{figure}

Due to high correlation among different measures of centrality, the comments are limited to the simplest degree centrality. In Figure \ref{figure:5} it is represented the degree distribution of scholars who serve as editorial leaders in at least one economics journal. This distribution is right-skewed, with 197 (6.8\%) isolated editorial leaders, 316 (11\%) linked to only one other editorial leader, and the majority having less than six links. On the far right tail, there are 21 editorial leaders linked to over 60 other editorial leaders. Table \ref{table:21} presents the top ten of the editorial leaders according to their degree.  

\begin{table}[ht!]
\caption{Editorial leaders with highest degree.}
\label{table:21}
\centering
\begin{tabular}{|l|cc|}
\hline
\textbf{Name} & \textbf{Degree} & \textbf{Rank Degree} \\
\hline
James J. Heckman & 103 & 1 \\
Douglas J. Cumming & 82 & 2 \\
Brian M. Lucey & 81 & 3 \\
Vernon L. Smith & 79 & 4 \\
Keun Lee & 77 & 5 \\
Thanasis Stengos & 75 & 6 \\
Stephen J. Turnovsky & 75 & 6 \\
Menzie David Chinn & 73 & 8 \\
Andrés Rodríguez-Pose & 72 & 9 \\
Oliver E. Williamson & 70 & 10 \\
\hline
\end{tabular}
\end{table}

James J.Heckman  holds the top position; he serves as Editor-in-Chief for one of the Top Five Journals, the \emph{Journal of Political Economy}, also sits on 12 other editorial boards, including the \emph{Economics: The Open-Access, Open-Assessment E-Journal}, which was previously identified in Table \ref{table:18} as the most central journal in the interlocking editorship network. Douglas J. Cumming, who ranks second, is the Editor-in-Chief of \emph{Annals of Corporate Governance}, sits also in other 17 editorial boards, including the \emph{Journal of Risk and Financial Management}, which is the second most central journal, and the \emph{Journal of International Business Studies}, which is the eight most central journal. Brian M. Lucey, who serves as Editor-in-Chief for the \emph{International Review of Economics and Finance}, sits also in other 12 editorial boards, including the \emph{Journal of Risk and Financial Management}, \emph{Panoeconomicus}, \emph{Emerging Markets Finance and Trade}, which are among the top ten most central journals. All the remaining most central editorial leaders sits in at least one board of most central journals.  

These results seem to indicate that the `strategy' we have observed journals adopt to gain more prestige, selecting scholars who serve as editorial leader for other journals as members of their editorial boards, not only gives centrality to these journals, but also to the scholars who accept to sit in their boards. Thus, it is challenging to separate the editorial power from the scholarly prestige of individuals who serve as editorial leaders. Indeed, if a scholar has editorial power by serving as editorial leader in a journal, then they can enhance their individual prestige by accepting invitations to sit on the editorial boards of other journals seeking to bolster their own prestige by selecting renowned scholars. 

%\begin{table}[ht!]
%\caption{Correlation among the centrality rankings of the network of EIC}
%\label{table:21a}
%\centering
%\begin{tabular}{|l|c|c|c|c|}
%\hline
%& \textbf{All degree rank} & \textbf{Betweenness rank} &
%\textbf{Closeness rank} \\
%\hline
%\textbf{All degree rank} & 1 & 0.821 & 0.835 \\
%\textbf{Betweenness rank} & & 1 & 0.875 \\
%\textbf{Closeness rank} & & & 1 \\
%\hline
%\end{tabular}
%\end{table}

\begin{figure}[ht]
\centering
\includegraphics{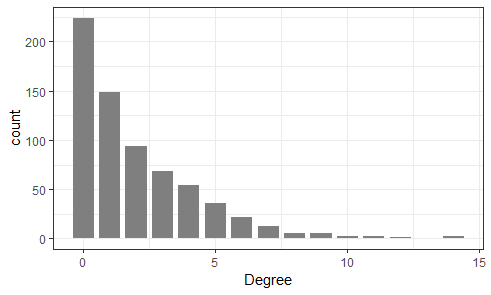}
\caption{Degree distribution of female scholars who serve as editorial leader in at least one economics journal. "Degree" refers to the number of links to other female scholars who also serve as editorial leader in at least one economics journal.}
\label{figure:6}
\end{figure}

As all the most central editorial leaders are male, a focus is made on the network generated by female editorial leaders. 
In this case as well, the network formed by female editorial leaders exhibits very different characteristics, as women editors generally tend to hold fewer seats simultaneously. The degree distribution, shown in Figure \ref{figure:6},  reveals that a significant proportion of female editorial leaders, 32.89\% (224 out of 705), are not linked to other female editorial leaders. The average degree is only 2. Table \ref{table:22} lists the scholars with the highest degrees. The maximum degree is 14 held by Jennifer L. Blouin and Judith Clifton.  

Jennifer L. Blouin, who serves as Editor-in-Chief for the \emph{Review of Accounting Studies}, sits in 6 other boards; analogously Judith Clifton, who serves as Editor-in-Chief for the \emph{Cambridge Journal of Regions, Economy and Society}, sits in 5 other boards. In both cases, none of the journals where they sit are among the most central journals in the network of journals and in the network of journals based on female editors.

\begin{table}[ht!]
\caption{Female editorial leaders with highest degree.}
\label{table:22}
\centering
\begin{tabular}{|l|cc|l|cc|}
\hline
\textbf{Name} & \textbf{Degree} & \textbf{Rank} & \textbf{Name} & \textbf{Degree} & \textbf{Rank} \\
\hline
Jennifer L. Blouin & 14 & 1 & Elisa Giuliani & 10 & 7 \\
Judith Clifton & 14 & 1 & Renee Adams & 9 & 10 \\
Sebnem Kalemli-Özcan & 12 & 3 & Catherine Tucker & 9 & 10 \\
Leah Boustan & 11 & 4 & Amy K. Glasmeier & 9 & 10 \\
Emmanuelle Auriol & 11 & 4 & Xuan Tian & 9 & 10 \\
Lori A. Beaman & 11 & 4 & Diane W. Schanzenbach & 9 & 10 \\
Mar Reguant & 10 & 7 & Ping Wang & 9 & 10 \\
\hline
\end{tabular}
\end{table}

These results seem to confirm that women tend to participate less to the `strategic decisions' in the selection of the editorial board members. Women are often less invited to join editorial boards of journals with the goal of enhancing the publication's prestige. This may be due to the challenge women face in translating their editorial power as editorial leaders into personal prestige, particularly in a male-dominated discipline as economics.

\section{Conclusions and policy recommendations}
\label{Conclusion}

Editorial board members have been identified as the gatekeepers of scientific knowledge, as they determine which research is published and their decisions significantly impact individual careers. Consequently, editorial roles, especially in leadership positions within highly influential journals, hold considerable power over the discipline. This study conducted an exploratory analysis of the composition of editorial boards for 1,516 active economics journals in 2019. The analysis examined the individual characteristics of board members and the main features of the interlocking editorship networks they generated.

The editorial boards of economics journals display a high degree of homophily, meaning that members tend to share similar characteristics. This phenomenon can be observed in various ways. Firstly, the boards member come in majority from Unites States institutions and are mainly men. In fact, more than the 33\% of all seats and more than the 35\% of editorial leader seats are held by scholars affiliated in the United States. The geographic concentration is higher in editorial leader seats, whose scholars come from 84 countries, compared to 151 countries declared for all editorial roles. Moreover, the 5 most represented countries hold the 42\% of all editorial seats and the 49\% of editorial leader  seats. In comparison with the proportions of country  affiliation of economics as registered in \citet{repec2023}, the United States, the United Kingdom, Canada, Australia, Turkey and Netherlands are over-represented in the editorial boards, while all the other countries are under-represented. In particular, China and Russia are not among the top 10 countries most represented on editorial boards but they are among the top 10 countries most represented in terms of the proportion of economics authors. Furthermore, 33\% of journals have over 50\% of their members affiliated with the same country, which is the United States in most cases. This editorial board members are mainly affiliated to élite universities, in particular from the United States and United Kingdom. For editorial leader seats the members come only from 1,036 institutions (the 17\% of total affiliations of our database). The gender composition of editorial boards reflect the under-representation of women in economics and, unlike in academic positions, suggests the existence of a a form of ‘horizontal segregation' of women in certain editorial boards. Indeed, women hold about 24\% of the available seats in the editorial boards and in the editorial leader seats, which is a proportion similar to that of female authors in economics \citep{repec2023}. However, a very high proportion (87\%) of editorial boards are composed of more than 50\% of male scholars. 

Secondly, we have developed a network analysis, which mainly consisted in comparing three different journal networks: the complete interlocking editorship network, generated by the crossed presence in different boards of the same scholars; the No-EL network, generated by scholars who do not hold an editorial leader position; and the EL network generated by scholars who hold at least one editorial leader position. The three networks shows very high distance correlations: the information they contain are very similar. Moreover, the clusters of journals identified in the three networks are also highly associated. 
These results suggest the existence of a high degree of social and intellectual homophily inside each board: different members in different roles tend to generate similar connections between journals. This may be due to the indirect role of Editors-in-Chief in selecting as editorial board members scholars ‘similar’ to them. 

Finally, the analysis of the most central journals and scholars in the networks suggests that journals probably adopt `strategic decisions' in the selection of the editorial board members. Some of the most central journals appear to select as members of their editorial boards scholars who serve as editorial leader for other journals. This enhances the centrality in the network of these journals, but also of the scholars who accept to sit in their boards. It can be speculated that this reinforcement mechanism is designed to trigger an editorial Matthew effect, which involves both journals and scholars simultaneously. The mechanism could translate the editorial power owned by an editorial leader in personal prestige; in turn the presence of prestigious scholars in a board may reinforce their prestige and possibly the editorial power of editorial leaders. In view of corroborating this conjecture future research should explore the relation between centrality in the interlocking editorship network and measures of impact of journals.
Data showed that women tend to be excluded from this editorial game. 

The results obtained suggest that economics journals, and hence economics as a field, are characterized by editorial boards that are dominated by scholars from the United States, with a prevalence of men and a high concentration of editorial power among a few élite institutions and scholars that are socially and intellectually similar. The strategic selection of board members seems to reinforce this homophily. As \citet{hodgson1999} have already warned, this high concentration of editorial power carries a serious risk for innovative research in economics. Therefore, it is crucial to implement practices aimed at minimizing this concentration in order to foster pluralism.

The study we have presented is helpful in achieving this objective as it not only confirms the findings of other partial studies on the same topic but also broadens the concept of homophily that editors should consider when selecting their board members. Our analysis also reveals that some journals published by scholarly societies that prioritize gender diversity representation, for example, exhibit a higher female presence than others. Thus, it is likely that some steps in this direction are being taken. However, as this analysis shows, these measures are not sufficient and likely rely on only a few factors. In our opinion, efforts should be directed towards reducing homophily in a broader sense. In this regard, the Guidelines for New Editorial Appointments of the American Economic Association \citet{AEA} introduce a valuable innovation: they emphasize that ‘‘Editors are encouraged to consider how a candidate would add to the diversity of the existing board, including (but not limited to) intellectual diversity (methods and fields of study), institutional diversity (where a person works and where they were trained), demographic diversity (including gender, race and ethnicity), and geographic diversity (national and international)". In addition to this prescription, there are term limits for editorial board members, particularly in editorial leader positions, that are implemented to prevent the long-term concentration of power. These updated procedures seem a very interesting innovation to promote diversity, ensure regular turnover, and include fresh perspectives, and they probably should be adopted by all journals.

\newpage
\section*{APPENDIX}\label{Appendix A}
\renewcommand{\thesection}{A\arabic{section}}

For the affiliations, we have focused our attention on what was declared on the websites of the journals. In the case of the University of California, the campus was not always specified, making it challenging to uniformly determine which campus the scholars belong to, as the majority simply indicated `University of California' generically. Therefore, we will present the distribution of seat affiliations as declared.

\begin{table}[ht!]
\setcounter{table}{0}
\renewcommand{\thetable}{A\arabic{table}}
\caption{University of California seats distribution}
\label{table:A}
\centering
\begin{tabular}{|l|cc|}
\hline
& \textbf{All Editorial Roles} & \textbf{Editorial Leaders} \\
\hline
University of California & 498 & 35 \\
University of California Berkeley & 206 & 11 \\
University of California Los Angeles & 132 & 8 \\
University of California Davis & 71 & 6 \\
University of California Irvine & 65 & 5 \\
University of California Riverside & 47 & 4 \\
University of California Santa Barbara & 36 & 1 \\
University of California Santa Cruz & 27 & 1 \\
University of California San Francisco & 3 & \\
\hline
\end{tabular}
\end{table}

\bibliographystyle{abbrvnat}
\bibliography{references}  %%% Uncomment this line and comment out the ``thebibliography'' section below to use the external .bib file (using bibtex) .

\begin{thebibliography}{42}
\providecommand{\natexlab}[1]{#1}
\providecommand{\url}[1]{\texttt{#1}}
\expandafter\ifx\csname urlstyle\endcsname\relax
  \providecommand{\doi}[1]{doi: #1}\else
  \providecommand{\doi}{doi: \begingroup \urlstyle{rm}\Url}\fi

\bibitem[Addis and Villa(2003)]{addis2003}
E.~Addis and P.~Villa.
\newblock The editorial boards of italian economics journals: women, gender, and social networking.
\newblock \emph{Feminist Economics}, 9\penalty0 (1):\penalty0 75--91, 2003.
\newblock \doi{10.1080/1354570032000057062}.
\newblock URL \url{https://www.tandfonline.com/doi/abs/10.1080/1354570032000057062}.

\bibitem[{American Economic Association}(2022)]{AEA}
{American Economic Association}.
\newblock Aea guidelines for new editorial appointments, 2022.
\newblock URL \url{https://www.aeaweb.org/journals/policies/appointment-procedures#diversity}.

\bibitem[Andrikopoulos and Economou(2015)]{andrikopoulos2015}
A.~Andrikopoulos and L.~Economou.
\newblock Editorial board interlocks in financial economics.
\newblock \emph{International review of financial analysis}, 37:\penalty0 51--62, 2015.
\newblock \doi{10.1016/j.irfa.2014.11.015}.
\newblock URL \url{https://www.sciencedirect.com/science/article/pii/S1057521914001896?casa_token=J_ULr_doZb0AAAAA:4pDfHwcG4hUv2j_lsybwK8Yli0RVqH1-Uqr-fG3t-6NGAwxAhznhkt1NwR_LIIIzJn_jcNOn}.

\bibitem[Baccini(2009)]{baccini2009}
A.~Baccini.
\newblock Italian economic journals. a network-based ranking and an exploratory analysis of their influence on setting international professional standards.
\newblock \emph{Rivista italiana degli economisti}, 14\penalty0 (3):\penalty0 491--512, 2009.
\newblock \doi{10.1427/31429}.
\newblock URL \url{https://www.rivisteweb.it/doi/10.1427/31429}.

\bibitem[Baccini and Barabesi(2010)]{baccini2010}
A.~Baccini and L.~Barabesi.
\newblock Interlocking editorship. a network analysis of the links between economic journals.
\newblock \emph{Scientometrics}, 82\penalty0 (2):\penalty0 365--389, 2010.
\newblock \doi{10.1007/s11192-009-0053-7}.
\newblock URL \url{https://akjournals.com/view/journals/11192/82/2/article-p365.xml}.

\bibitem[Baccini and Barabesi(2011)]{baccini2011}
A.~Baccini and L.~Barabesi.
\newblock Seats at the table: The network of the editorial boards in information and library science.
\newblock \emph{Journal of informetrics}, 5\penalty0 (3):\penalty0 382--391, 2011.
\newblock \doi{10.1016/j.joi.2011.01.012}.
\newblock URL \url{https://www.sciencedirect.com/science/article/pii/S1751157711000137?casa_token=q0wb5YUtSFwAAAAA:-GH_ktQiE7HUgJKL0MKhaZ1K50cTzw3oI4qcgYDebixyk1ls7UMax-hdBu3afbJFh5vhj61I}.

\bibitem[Baccini et~al.(2009)Baccini, Barabesi, and Marcheselli]{bacciniet2009}
A.~Baccini, L.~Barabesi, and M.~Marcheselli.
\newblock How are statistical journals linked? a network analysis.
\newblock \emph{Chance}, 22\penalty0 (3):\penalty0 35--45, 2009.
\newblock \doi{10.1080/09332480.2009.10722969}.
\newblock URL \url{https://www.tandfonline.com/doi/abs/10.1080/09332480.2009.10722969?journalCode=ucha20}.

\bibitem[Baccini et~al.(2020)Baccini, Barabesi, Khelfaoui, and Gingras]{baccini2020}
A.~Baccini, L.~Barabesi, M.~Khelfaoui, and Y.~Gingras.
\newblock Intellectual and social similarity among scholarly journals: An exploratory comparison of the networks of editors, authors and co-citations.
\newblock \emph{Quantitative Science Studies}, 1\penalty0 (1):\penalty0 277--289, 2020.
\newblock \doi{10.1162/qss_a_00006}.
\newblock URL \url{https://direct.mit.edu/qss/article/1/1/277/15560/Intellectual-and-social-similarity-among-scholarly}.

\bibitem[Bastian et~al.(2009)Bastian, Heymann, and Jacomy]{bastian2009}
M.~Bastian, S.~Heymann, and M.~Jacomy.
\newblock Gephi: an open source software for exploring and manipulating networks.
\newblock \emph{Proceedings of the international AAAI conference on web and social media}, 3:\penalty0 361--362, 2009.
\newblock \doi{10.1609/icwsm.v3i1.13937}.
\newblock URL \url{https://ojs.aaai.org/index.php/ICWSM/article/view/13937}.

\bibitem[Blondel et~al.(2008)Blondel, Guillaume, Lambiotte, and Lefebvre]{Louvain}
V.~D. Blondel, J.-L. Guillaume, R.~Lambiotte, and E.~Lefebvre.
\newblock Fast unfolding of communities in large networks.
\newblock \emph{Journal of Statistical Mechanics: Theory and Experiment}, 2008\penalty0 (10):\penalty0 P10008, oct 2008.
\newblock \doi{10.1088/1742-5468/2008/10/p10008}.
\newblock URL \url{https://arxiv.org/abs/0803.0476}.

\bibitem[Braun and Di{\'o}spatonyi(2005{\natexlab{a}})]{braun2005a}
T.~Braun and I.~Di{\'o}spatonyi.
\newblock Counting the gatekeepers of international science journals a worthwhile science indicator.
\newblock \emph{Current Science}, 89\penalty0 (9):\penalty0 1548--1551, 2005{\natexlab{a}}.
\newblock \doi{10.3103/S0147688216030035}.
\newblock URL \url{http://www.jstor.org/stable/24110926}.

\bibitem[Braun and Di{\'o}spatonyi(2005{\natexlab{b}})]{braun2005b}
T.~Braun and I.~Di{\'o}spatonyi.
\newblock World flash on basic research.
\newblock \emph{Scientometrics}, 62\penalty0 (3):\penalty0 297--319, 2005{\natexlab{b}}.
\newblock \doi{10.1007/s11192-005-0023-7}.
\newblock URL \url{https://link.springer.com/article/10.1007/s11192-005-0023-7}.

\bibitem[Crane(1967)]{crane1967}
D.~Crane.
\newblock The gatekeepers of science: Some factors affecting the selection of articles for scientific journals.
\newblock \emph{The American Sociologist}, pages 195--201, 1967.
\newblock URL \url{https://www.jstor.org/stable/27701277}.

\bibitem[Csom{\'o}s and Lengyel(2022)]{csomos2022}
G.~Csom{\'o}s and B.~Lengyel.
\newblock Geographies of the global co-editor network in oncology.
\newblock \emph{PloS one}, 17\penalty0 (3):\penalty0 e0265652, 2022.
\newblock \doi{10.1371/journal.pone.0265652}.
\newblock URL \url{https://journals.plos.org/plosone/article?id=10.1371/journal.pone.0265652}.

\bibitem[De~Grazia(1963)]{degrazia1963}
A.~De~Grazia.
\newblock The scientific reception system and dr. velikovsky.
\newblock \emph{American Behavioral Scientist}, 7\penalty0 (1):\penalty0 45--49, 1963.
\newblock \doi{10.1177/000276426300700106}.
\newblock URL \url{https://journals.sagepub.com/doi/abs/10.1177/000276426300700106?journalCode=absb}.

\bibitem[Duarte and Giraud(2016)]{duarte2016}
P.~G. Duarte and Y.~Giraud.
\newblock The place of the history of economic thought in mainstream economics, 1991--2011, viewed through a bibliographic survey.
\newblock \emph{Journal of the history of economic thought}, 38\penalty0 (4):\penalty0 431--462, 2016.
\newblock \doi{10.1017/S1053837216000481}.

\bibitem[Ductor and Visser(2023)]{ductor2023}
L.~Ductor and B.~Visser.
\newblock Concentration of power at the editorial boards of economics journals.
\newblock \emph{Journal of Economic Surveys}, 37\penalty0 (2):\penalty0 189--238, 2023.
\newblock \doi{10.1111/joes.12497}.
\newblock URL \url{https://onlinelibrary.wiley.com/doi/full/10.1111/joes.12497}.

\bibitem[Faria(2005)]{faria2005}
J.~R. Faria.
\newblock The game academics play: Editors versus authors.
\newblock \emph{Bulletin of Economic research}, 57\penalty0 (1):\penalty0 1--12, 2005.
\newblock \doi{10.1111/j.1467-8586.2005.00212.x}.
\newblock URL \url{https://onlinelibrary.wiley.com/doi/10.1111/j.1467-8586.2005.00212.x}.

\bibitem[Gibbons and Fish(1991)]{gibbons1991}
J.~D. Gibbons and M.~Fish.
\newblock Rankings of economics faculties and representation on editorial boards of top journals.
\newblock \emph{The Journal of Economic Education}, 22\penalty0 (4):\penalty0 361--372, 1991.
\newblock \doi{10.1080/00220485.1991.10844728}.
\newblock URL \url{https://www.tandfonline.com/doi/abs/10.1080/00220485.1991.10844728}.

\bibitem[Goyanes and De-Marcos(2020)]{goyanes2020}
M.~Goyanes and L.~De-Marcos.
\newblock Academic influence and invisible colleges through editorial board interlocking in communication sciences: a social network analysis of leading journals.
\newblock \emph{Scientometrics}, 123\penalty0 (2):\penalty0 791--811, 2020.
\newblock \doi{10.1007/s11192-020-03401-z}.
\newblock URL \url{https://link.springer.com/article/10.1007/s11192-020-03401-z}.

\bibitem[Hatfield et~al.(1995)Hatfield, Ostbye, and Sori]{hatfield1995}
C.~Hatfield, T.~Ostbye, and C.~Sori.
\newblock Sex of editor in medical journals.
\newblock \emph{The Lancet}, 8950\penalty0 (345):\penalty0 662, 1995.
\newblock \doi{10.1016/s0140-6736(95)90572-3}.
\newblock URL \url{https://www.thelancet.com/pdfs/journals/lancet/PIIS0140-6736(95)90572-3.pdf}.

\bibitem[Hodgson and Rothman(1999)]{hodgson1999}
G.~M. Hodgson and H.~Rothman.
\newblock The editors and authors of economics journals: A case of institutional oligopoly?
\newblock \emph{The economic journal}, 109\penalty0 (453):\penalty0 165--186, 1999.
\newblock \doi{10.1111/1468-0297.00407}.
\newblock URL \url{https://academic.oup.com/ej/article-abstract/109/453/165/5128687}.

\bibitem[Leydesdorff and Wagner(2009)]{leydesdorff2009}
L.~Leydesdorff and C.~Wagner.
\newblock Is the united states losing ground in science? a global perspective on the world science system.
\newblock \emph{Scientometrics}, 78\penalty0 (1):\penalty0 23--36, 2009.
\newblock \doi{10.1007/s11192-008-1830-4}.
\newblock URL \url{https://akjournals.com/view/journals/11192/78/1/article-p23.xml}.

\bibitem[Liwei and Chunlin(2015)]{liwei2015}
Z.~Liwei and J.~Chunlin.
\newblock Social network analysis and academic performance of the editorial board members for journals of library and information science.
\newblock \emph{COLLNET Journal of Scientometrics and Information Management}, 9\penalty0 (2):\penalty0 131--143, 2015.
\newblock \doi{10.1080/09737766.2015.1069947}.
\newblock URL \url{https://www.tandfonline.com/doi/abs/10.1080/09737766.2015.1069947}.

\bibitem[Lockstone-Binney et~al.(2021)Lockstone-Binney, Ong, and Mair]{lockstone2021}
L.~Lockstone-Binney, F.~Ong, and J.~Mair.
\newblock Examining the interlocking of tourism editorial boards.
\newblock \emph{Tourism Management Perspectives}, 38:\penalty0 100829, 2021.
\newblock \doi{10.1016/j.tmp.2021.100829}.
\newblock URL \url{https://www.sciencedirect.com/science/article/pii/S2211973621000428}.

\bibitem[Lundberg and Stearns(2019)]{lundberg2019}
S.~Lundberg and J.~Stearns.
\newblock Women in economics: Stalled progress.
\newblock \emph{Journal of Economic Perspectives}, 33\penalty0 (1):\penalty0 3--22, 2019.
\newblock \doi{10.1257/jep.33.1.3}.
\newblock URL \url{https://pubs.aeaweb.org/doi/pdfplus/10.1257/jep.33.1.3}.

\bibitem[Marcuzzo and Zacchia(2016)]{marcuzzo2016}
M.~C. Marcuzzo and G.~Zacchia.
\newblock Is history of economics what historians of economic thought do?: A quantitative investigation.
\newblock \emph{Is History of Economics What Historians of Economic Thought Do?: A Quantitative Investigation}, pages 29--46, 2016.
\newblock \doi{10.19272/201606103002}.
\newblock URL \url{http://digital.casalini.it/10.19272/201606103002}.

\bibitem[Maule{\'o}n et~al.(2013)Maule{\'o}n, Hill{\'a}n, Moreno, G{\'o}mez, and Bordons]{mauleon2013}
E.~Maule{\'o}n, L.~Hill{\'a}n, L.~Moreno, I.~G{\'o}mez, and M.~Bordons.
\newblock Assessing gender balance among journal authors and editorial board members.
\newblock \emph{Scientometrics}, 95\penalty0 (1):\penalty0 87--114, 2013.
\newblock \doi{10.1007/s11192-012-0824-4}.
\newblock URL \url{https://link.springer.com/article/10.1007/s11192-012-0824-4}.

\bibitem[Mazov and Gureev(2016)]{mazov2016}
N.~A. Mazov and V.~N. Gureev.
\newblock The editorial boards of scientific journals as a subject of scientometric research: a literature review.
\newblock \emph{Scientific and Technical Information Processing}, 43\penalty0 (3):\penalty0 144--153, 2016.
\newblock \doi{10.3103/S0147688216030035}.
\newblock URL \url{https://link.springer.com/article/10.3103/S0147688216030035}.

\bibitem[Merton(1942)]{merton1942}
R.~K. Merton.
\newblock A note on science and democracy.
\newblock \emph{Journal of Legal and Political Sociology}, 1:\penalty0 115--126, 1942.
\newblock URL \url{https://heinonline.org/HOL/Page?handle=hein.journals/jolegpo1&id=115&collection=journals}.

\bibitem[Metz et~al.(2016)Metz, Harzing, and Zyphur]{metz2016}
I.~Metz, A.-W. Harzing, and M.~J. Zyphur.
\newblock Of journal editors and editorial boards: who are the trailblazers in increasing editorial board gender equality?
\newblock \emph{British journal of management}, 27\penalty0 (4):\penalty0 712--726, 2016.
\newblock \doi{10.1111/1467-8551.12133}.
\newblock URL \url{https://onlinelibrary.wiley.com/doi/full/10.1111/1467-8551.12133?casa_token=O0fkOl_nhaIAAAAA%3ACoJFwA8aK0zGaeGOV2NJwoh9K2M2tcl0aM9KqTrK0_f_6dFvqYSHPXdxXwp4An4da9IqfAvYKEnl}.

\bibitem[Ni and Ding(2010)]{ni2010}
C.~Ni and Y.~Ding.
\newblock Journal clustering through interlocking editorship information.
\newblock \emph{Proceedings of the American Society for Information Science and Technology}, 47\penalty0 (1):\penalty0 1--10, 2010.
\newblock \doi{10.1002/meet.14504701202}.
\newblock URL \url{https://asistdl.onlinelibrary.wiley.com/doi/full/10.1002/meet.14504701202}.

\bibitem[{R Core Team}(2013)]{R_soft}
{R Core Team}.
\newblock R: A language and environment for statistical computing, 2013.
\newblock URL \url{http://www.R-project.org/}.

\bibitem[Rand(1971)]{Rand}
W.~M. Rand.
\newblock Objective criteria for the evaluation of clustering methods.
\newblock \emph{Journal of the American Statistical Association}, 66\penalty0 (336):\penalty0 846--850, 1971.
\newblock \doi{10.1080/01621459.1971.10482356}.

\bibitem[RePEc(2023)]{repec2023}
RePEc.
\newblock Female representation in economics, as of february 2023, 2023.
\newblock URL \url{https://ideas.repec.org/top/female.html}.

\bibitem[Stegmaier et~al.(2011)Stegmaier, Palmer, and Van~Assendelft]{stegmaier2011}
M.~Stegmaier, B.~Palmer, and L.~Van~Assendelft.
\newblock Getting on the board: the presence of women in political science journal editorial positions.
\newblock \emph{PS: Political science \& politics}, 44\penalty0 (4):\penalty0 799--804, 2011.
\newblock \doi{10.1017/S1049096511001284}.

\bibitem[Székely and Rizzo(2014)]{székely2014}
G.~J. Székely and M.~L. Rizzo.
\newblock Partial distance correlation with methods for dissimilarities.
\newblock \emph{Annals of Statistics}, 42\penalty0 (6):\penalty0 2382--2412, 12 2014.
\newblock \doi{10.1214/14-AOS1255}.
\newblock URL \url{https://projecteuclid.org/journals/annals-of-statistics/volume-42/issue-6/Partial-distance-correlation-with-methods-for-dissimilarities/10.1214/14-AOS1255.full}.

\bibitem[Székely et~al.(2007)Székely, Rizzo, and Bakirov]{szekely2007}
G.~J. Székely, M.~L. Rizzo, and N.~K. Bakirov.
\newblock Measuring and testing dependence by correlation of distances.
\newblock \emph{The Annals of Statistics}, 35\penalty0 (6):\penalty0 2769--2794, 2007.
\newblock \doi{10.1214/009053607000000505}.
\newblock URL \url{https://projecteuclid.org/journals/annals-of-statistics/volume-35/issue-6/Measuring-and-testing-dependence-by-correlation-of-distances/10.1214/009053607000000505.full}.

\bibitem[Teixeira and Oliveira(2018)]{teixeira2018}
E.~K. Teixeira and M.~Oliveira.
\newblock Editorial board interlocking in knowledge management and intellectual capital research field.
\newblock \emph{Scientometrics}, 117\penalty0 (3):\penalty0 1853--1869, 2018.
\newblock \doi{10.1007/s11192-018-2937-x}.
\newblock URL \url{https://link.springer.com/article/10.1007/s11192-018-2937-x}.

\bibitem[Traag et~al.(2019)Traag, Waltman, and van Eck]{Traag}
V.~A. Traag, L.~Waltman, and N.~J. van Eck.
\newblock From louvain to leiden: guaranteeing well-connected communities.
\newblock \emph{Scientific Reports}, 9\penalty0 (1):\penalty0 5233, 2019.
\newblock ISSN 2045-2322.
\newblock \doi{10.1038/s41598-019-41695-z}.

\bibitem[Wu et~al.(2020)Wu, Lu, Li, and Li]{wu2020}
D.~Wu, X.~Lu, J.~Li, and J.~Li.
\newblock Does the institutional diversity of editorial boards increase journal quality? the case economics field.
\newblock \emph{Scientometrics}, 124\penalty0 (2):\penalty0 1579--1597, 2020.
\newblock \doi{10.1007/s11192-020-03505-6}.
\newblock URL \url{https://link.springer.com/article/10.1007/s11192-020-03505-6}.

\bibitem[Zsindely et~al.(1982)Zsindely, Schubert, and Braun]{zsindely1982}
S.~Zsindely, A.~Schubert, and T.~Braun.
\newblock Editorial gatekeeping patterns in international science journals. a new science indicator.
\newblock \emph{Scientometrics}, 4\penalty0 (1):\penalty0 57--68, 1982.
\newblock \doi{10.1007/bf02098006}.
\newblock URL \url{https://akjournals.com/view/journals/11192/4/1/article-p57.xml}.

\end{thebibliography}





% \end{thebibliography}

\end{document}